\documentclass[12pt]{spieman}  
\usepackage{tocloft}
\usepackage{amsmath,amsfonts,amssymb}
\usepackage{mathabx}
\usepackage{deluxetable}
\usepackage{graphicx}
\usepackage{setspace}
\usepackage{float}

\def\gappeq{\mathrel{ \rlap{\raise.5ex\hbox{$>$}}
                      {\lower.5ex\hbox{$\sim$}}  } }

\def\starketal{Stark et al.}
\def\savetal{Savransky et al.}

\title{Maximized ExoEarth Candidate Yields for Starshades}

\author[a]{Christopher C. Stark}
\author[b]{Stuart Shaklan}
\author[b]{Doug Lisman}
\author[b]{Eric Cady}
\author[c]{Dmitry Savransky}
\author[d]{Aki Roberge}
\author[d]{Avi M. Mandell}

\affil[a]{Space Telescope Science Institute, 3700 San Martin Dr, Baltimore, MD 21218, USA; cstark@stsci.edu}
\affil[b]{Jet Propulsion Laboratory, California Institute of Technology, Pasadena, CA 91109, USA}
\affil[c]{Sibley School of Mechanical and Aerospace Engineering, Cornell University, Ithaca, NY 14853, USA}
\affil[d]{NASA Goddard Space Flight Center, Greenbelt, MD 20771, USA}

\cftpagenumbersoff{figure}
\cftpagenumbersoff{table} 
\begin{document} 
\maketitle

\begin{abstract}
The design and scale of a future mission to directly image and characterize potentially Earth-like planets will be impacted, to some degree, by the expected yield of such planets.  Recent efforts to increase the estimated yields, by creating observation plans optimized for the detection and characterization of Earth-twins, have focused solely on coronagraphic instruments; starshade-based missions could benefit from a similar analysis.  Here we explore how to prioritize observations for a starshade given the limiting resources of both fuel and time, present analytic expressions to estimate fuel use, and provide efficient numerical techniques for maximizing the yield of starshades.  We implemented these techniques to create an approximate design reference mission code for starshades and used this code to investigate how exoEarth candidate yield responds to changes in mission, instrument, and astrophysical parameters for missions with a single starshade.  We find that a starshade mission operates most efficiently somewhere between the fuel- and exposure-time limited regimes, and as a result, is less sensitive to photometric noise sources as well as parameters controlling the photon collection rate in comparison to a coronagraph.  We produced optimistic yield curves for starshades, assuming our optimized observation plans are schedulable and future starshades are not thrust-limited.  Given these yield curves, detecting and characterizing several dozen exoEarth candidates requires either multiple starshades or an $\eta_{\rm Earth}\gtrsim0.3$.
\end{abstract}


{\noindent \footnotesize\textbf{*}Christopher C. Stark,  \linkable{cstark@stsci.edu} }

\begin{spacing}{2}   

\section{Introduction}
\label{intro}

The design of future large aperture space missions will be influenced by the desire to detect and characterize potentially Earth-like planets around other stars (exoEarth candidates).  Recent estimates of the yield of exoEarth candidates have focused primarily on missions that use an internal coronagraph to suppress starlight \cite{brown2005,brownsoummer2010,stark2014_2,stark2015}.  However, coronagraphs can have relatively low throughput, potentially limited bandwidth requiring many observations to obtain a full spectrum, and require extremely stable mirrors.  External occulters (starshades), on the other hand, have relatively high throughput, operate over a relatively wide bandpass, and because the starlight is suppressed prior to diffraction by the primary mirror, do not require similar levels of telescope mirror stability \cite{cash2006, vanderbei2007}.

Yield estimates for coronagraph-based mission concepts have evolved recently from noise floor-limited searches to Earth-twin optimized searches. These optimized searches select individual exposure times to maximize overall mission yield, resulting in roughly a tripling of the expected yields for coronagraphs \cite{stark2014_2,stark2015}.  Scaling relationships from these yield-maximized estimates have informed coronagraph-based mission design, e.g., yield is more sensitive to telescope aperture than total mission lifetime \cite{stark2014_2}.  While studies of starshade-based mission yields have explored the impact of astrophysical parameters and the starshade's flight path for specific mission concepts \cite{savransky2010,turnbull2012,trabert2015}, yield models have not yet investigated in detail the balance between exposure time and fuel use necessary to maximize the yield of a starshade-based mission.

Here we apply the optimization techniques developed by \starketal\cite{stark2014_2,stark2015} to starshade-based missions. We first solve for the optimal observation plan for the mission, then estimate the mission-long fuel requirements of such a plan using scaling relationships.  In Section \ref{caveats} we present our baseline mission assumptions.  We then discuss updates to the optimization code for the case of starshades in Section \ref{methods_section}.  We present and discuss our results in Sections \ref{results_section} and \ref{discussion_section}.

\section{Assumptions \& Caveats}
\label{caveats}

\subsection{Scope and Intent of Models}

ExoEarth candidate yield calculations are performed with what is commonly referred to as a design reference mission (DRM) code.  Yield estimates published in the literature have been calculated using different DRM codes that vary significantly in scope, complexity, and intent.  Some DRM codes are simply back-of-the-envelope calculations based on the exposure times of Earth twins at quadrature around nearby stars.  Other DRM codes explicitly calculate the visibility of every star given a well defined field of regard for the mission, create a realistic observation schedule, and perform a dynamic traveling salesman-type calculation.  For starshade-based missions, most yield estimates have adopted the latter, which we refer to as ``mission execution simulators" because they simulate a mission from beginning to end, making decisions on the fly and calculating starshade fuel used when slewing between specific targets \cite{trabert2015}.  

Our DRM code, like the code of Ref.~\citenum{stark2014_2} and \citenum{stark2015}, is neither a back-of-the-envelope calculation, nor a mission execution simulator.  Our code creates an optimized \emph{static} observation plan, which, when executed, will produce higher yields than any other static observation plan.  By static, we mean that the code does not update the exposure times of the plan based on information learned over the course of the mission.  However, our code inherently assumes that spectral characterization time is only applied to true exoEarth candidates, which can only be true if the decision to characterize is made on the fly.  Our code handles the bookkeeping for this spectral characterization time by weighting each observation's possible spectral characterization time by the probability that a planet is detected during that observation.  Thus, it may be more accurate to call our optimized observation plans semi-static.

Our code does not determine the absolute times or order of observations, and therefore does not determine an observation schedule.  Rather, we assume our observation plan is schedulable and that an efficient path is possible for the starshade.  Previous studies have shown that it is possible to schedule an efficient starshade path that meets fuel constraints for a predetermined set of targets and observations \cite{turnbull2012, trabert2015}.

The fundamentals of our exoEarth candidate yield calculations are based on the completeness methods pioneered by Ref.~\citenum{brown2005}.  We start by distributing a large number of synthetic exoEarths around known main sequence stars, drawn from the Hipparcos catalog and vetted for binarity \cite{stark2014_2}.  We distribute a cloud of synthetic planets around each star, sampling all possible habitable zone orbits, inclinations, and phases.  We illuminate each synthetic planet with starlight and calculate the planet's exposure time.  For each star, we can then determine the fraction of the synthetic cloud that is observable (i.e., completeness) as a function of exposure time.  The completeness is equivalent to the probability of detecting the planet, if that planet exists.

Based on the completeness of each star as a function of time, we then prioritize stars by a benefit-to-cost ratio and down-select to those observations that fit within the mission lifetime.  Our code can calculate multiple visits to each star, enabling the detection of planets that were unobservable during the previous visit.  Our code also optimally distributes exposure time, such that it maximizes yield \cite{stark2015}.  

Here we extend this code to starshades.  To facilitate a direct comparison with the coronagraph results of Ref.~\citenum{stark2015}, we make identical astrophysical assumptions where possible.  For a detailed description of the code and justification of assumptions, we refer the reader to Ref.~\citenum{stark2014_2} and \citenum{stark2015}.

\subsection{Baseline Astrophysical Assumptions\label{astro_assumptions_section}}

With exception to the local zodiacal light, we made the same astrophysical assumptions as \starketal\cite{stark2015}.  We adopted the same habitable zone (HZ), given by Ref.~\citenum{kopparapu2013} for a Sun-like star and scaling with stellar luminosity.  We assumed all synthetic exoEarths are on circular orbits, spaced logarithmically in semi-major axis, and have a V band geometric albedo of $0.2$.  We also adopted the same exozodi definition and baseline exozodi level of 3 zodis, the details of which are described in \starketal\cite{stark2014_2,stark2015}.  We assumed the same baseline value of $\eta_{\Earth}=0.1$ (except for Section \ref{savransky_comparison_section}, in which we temporarily adopted $\eta_{\Earth}=0.5$ to compare our yield with previous results). We also used the same input target catalog, vetted in an identical fashion.  Table \ref{exoearth_params_table} summarizes all of our baseline astrophysical assumptions. 

\begin{deluxetable}{ccl}
\tablewidth{0pt}
\footnotesize
\tablecaption{Baseline Astrophysical Parameters\label{exoearth_params_table}}
\tablehead{
\colhead{Parameter} & \colhead{Value} & \colhead{Description} \\
}
\startdata
$\eta_{\Earth}$ & $0.1$ & Fraction of Sun-like stars with an exoEarth candidate \\
$R_{\rm p}$ & $1$ $R_{\Earth}$ & Planet radius \\
$a$ & $[0.75,1.77]$ AU\tablenotemark{*} & Semi-major axis (uniform in $\log{a}$) \\
$e$ & $0$ & Eccentricity (circular orbits) \\
$\cos{i}$ & $[-1,1]$ & Cosine of inclination (uniform distribution) \\
$\omega$ & $[0,2\pi)$ & Argument of pericenter (uniform distribution) \\
$M$ & $[0,2\pi)$ & Mean anomaly (uniform distribution) \\
$\Phi$ & Lambertian & Phase function \\ 
$A_G$ & $0.2$ & Geometric albedo of planet at $0.55$ and $1$ $\mu$m \\
$z$ & 22.1 mag arcsec$^{-2}$\tablenotemark{\dag}  & Average V band surface brightness of zodiacal light \\
$x$ & 22 mag arcsec$^{-2}$\tablenotemark{\ddag}  & V band surface brightness of 1 zodi of exozodiacal dust \\
$n$ & $3$ & Number of zodis for all stars \\
\enddata
\vspace{-0.1in}
\tablenotetext{*}{$a$ given for a solar twin.  The habitable zone is scaled by $\sqrt{L_{\star}/L_{\Sun}}$ after calculating projected separation $s_{\rm p}$.}
\tablenotetext{\dag}{Zodiacal light is brighter for a starshade than a coronagraph---see Section \ref{astro_assumptions_section}. Varies with ecliptic latitude---see Ref.~\citenum{leinert1998}.}
\tablenotetext{\ddag}{For Solar twin. Varies with spectral type---see Appendix C in Ref.~\citenum{stark2014_2}.}
\end{deluxetable}

\starketal\cite{stark2014_2,stark2015} assumed targets could be observed at solar longitudes $\sim135^{\circ}$, where the zodiacal light is minimized and varies slowly with solar longitude.  The zodiacal brightness was calculated for each star using this solar longitude and the star's ecliptic latitude (see Appendix B in \citenum{stark2014_2}).  As a result, the mean local zodiacal light brightness $z=23$ mag arcsec$^{-2}$.

However, to avoid reflecting sunlight into the telescope, starshades observe at solar elongations $30^{\circ}\lesssim\alpha\lesssim90^{\circ}$, where the local zodi can be as much as 14 times brighter.  To calculate the local zodiacal light surface brightness for each star, we transformed the stellar equatorial coordinates to ecliptic coordinates.  Because our code does not ``schedule" each observation or calculate detailed starshade/telescope pointing, we assumed each target is observed at the median solar elongation $\alpha = 60^{\circ}$.  Combined with the ecliptic coordinates of each target star, this assumed elongation determines the solar longitude and ecliptic latitude of observation.  We then interpolated Table 17 in Ref.~\citenum{leinert1998} to calculate the local zodiacal light surface brightness for each of these pointings.  For our baseline starshade mission, we found a mean local zodi brightness of $z=22.1$ mag arcsec$^{-2}$, as listed in Table \ref{exoearth_params_table} above.  In practice, we used the local zodi brightness calculated individually for each target.

\subsection{Baseline Mission Assumptions}

Table \ref{baseline_params_table} lists the baseline mission parameters.  For comparison with \starketal\cite{stark2015}, we duplicated as many of the mission parameters as possible.  ExoEarth candidates were assumed to be detected at V band and characterized at a wavelength of 1 $\mu$m to search for water.  Spectral characterization time counted against the total exposure time budget.  Only ExoEarth candidates simultaneously observable at $0.55$ and $1.0$ $\mu$m counted toward the yield.  We assumed exoEarth candidates and other planets or background objects can be distinguished either by color, less demanding spectral information, or a simple orbital analysis, such that negligible time is used taking spectra of non-exoEarth candidates.  We required the same characterization S/N and spectral resolving power requirements as \starketal\cite{stark2015}.

Like \starketal\cite{stark2015}, we allowed stars to be observed multiple times and included revisit completeness.  However, because many targets will have limited visibility windows for a starshade-based mission and the time between visits to a star will likely be dictated by schedulability, which we do not model, we only allowed up to 5 visits per star and did not allow the time between visits to be optimized.  For each target star, we set the time between visits equal to the median orbital period of synthetic HZ planets around that star, allowing the synthetic planets to mostly lose phase coherence \cite{brownsoummer2010}.  These assumptions are ultimately unimportant and impacted the yield by $\lesssim10\%$, as optimization typically selects just one or two visits to each star.

Like \starketal\cite{stark2015}, we ignored the details of overheads.  Unlike a coronagraph, starshades will not require an extended wavefront control time.  Therefore, whereas \starketal\cite{stark2015} assumed 1 year of exposure time and 1 year of wavefront control/overheads, we could have justifiably adopted 2 years of total exposure time for a starshade mission.  As we will show later, though, this time is better spent slewing than observing.  Thus, we adopted a total mission lifetime of 5 years and allowed our code to optimize the balance between slew and exposure time.  We assumed telescope repointing and fine alignment of the starshade make up a negligibly small fraction of the total mission lifetime.  

Compared to \starketal\cite{stark2015}, we changed a number of instrument parameters to better reflect expected starshade performance.  First, we required the starshade to operate over the full bandpass of $0.5$--$1.0$ $\mu$m.  Therefore, we adopted a contrast of $10^{-10}$ for both detection and characterization.  Because the identification of exoEarth candidates may require color information, we adopted half the starshade's bandpass as the bandwidth, or $0.22$ $\mu$m, which is twice as large as was assumed for the coronagraph in \starketal\cite{stark2015}.  Finally, we increased the total system throughput from $0.2$ to $0.65$, consistent with the optical throughput and detector quantum efficiency assumed by the Exo-S instrument design \cite{exos2015} (the usable fraction of the planet's light, $\Upsilon$, contained within the assumed photometric aperture is an additional factor that further decreases the effective throughput).

Unlike the coronagraph's inner working angle (IWA), expressed in units of $\lambda/D$ by \starketal\cite{stark2015}, we adopted a baseline 60 mas inner working angle (IWA) for our starshade that is independent of aperture size, and adopted an infinite OWA.  We reason that the IWA of a starshade is determined by the distance between the telescope and starshade as well as the diameter of the starshade, and will likely be limited by fuel usage and manufacturability.  We therefore should not expect the same scaling relationship between yield and aperture size as found by \starketal\cite{stark2015} for a coronagraph.

Together, the contrast, bandwidth, IWA, and telescope aperture size will determine the starshade's diameter $D_{\rm ss}$, distance $z$, and mass $m_{\rm ss}$, which we discuss in the following section.  All starshade diameters discussed in this paper are tip-to-tip diameters, and IWA $= D_{\rm ss}/2z$.  We note that the baseline mission parameters discussed here correspond to a $61.7$ m tip-to-tip diameter starshade at 106 Mm with a dry mass of 2742 kg.

We also implemented fuel constraints in our simulations.  We adopted a high-efficiency electric propulsion system for slewing with a specific impulse of 3000 s and a chemical propulsion system for station keeping with a specific impulse of 300 s.  For realistic propulsion systems, thrust will likely be limited to $< 1$ N.  However, in an effort to investigate upper limits on the yields for starshade missions, we chose to work far from the thrust-limited regime and adopted a thrust of 10 N, an assumption we address in Section \ref{yields_section}.  Finally, we chose a total mass limit of 9800 kg, the Delta IV Heavy payload mass limit to the Sun-Earth L2 point \cite{rioux2015}.  We investigate other mass limits in Section \ref{yields_section}.

We reduced the baseline telescope diameter from the 10 m aperture assumed by \starketal\cite{stark2015} to 4 m.  We do not intend the baseline starshade mission's yield calculated here to be similar to the baseline coronagraph's yield calculated by \starketal\cite{stark2015}.  The assumed baseline values simply represent a reasonable starting point comparable to previous works.  Later we will determine what mission parameters are required to achieve a given yield.

We note that not all of the baseline parameters are independent quantities.  For example, if we were to deviate from the baseline telescope diameter of 4 m, while keeping the contrast and IWA fixed, the starshade diameter would have to change.   We mark dependent quantities with a $\dagger$ symbol in Table \ref{baseline_params_table}.

\begin{deluxetable}{ccl}
\tablewidth{0pt}
\footnotesize
\tablecaption{Baseline Mission Parameters\label{baseline_params_table}}
\tablehead{
\colhead{Parameter} & \colhead{Value} & \colhead{Description} \\
}
\startdata
$D$ & $4$ m & Telescope diameter \\
$I_{\rm sk}$ & $300$ s & Specific impulse of station keeping propellant (chemical)\\
$I_{\rm slew}$ & $3000$ s & Specific impulse of slew propellant (electric)\\
$\epsilon_{\rm sk}$ & $0.8$ & Efficiency of station keeping fuel use\\
$\epsilon_{\rm slew}$ & $0.91$\tablenotemark{a\dagger} & Efficiency of slew fuel use\\
$D_{\rm ss}$ & $61.7$ m\tablenotemark{b\dagger} & Diameter of starshade\\
$m_{\rm dry}$ & $2742$ kg\tablenotemark{c\dagger} & Dry mass of starshade spacecraft including contingency\\
$m_{\rm tot}$ & $9800$ kg & Total initial mass of starshade spacecraft including fuel\\
$\mathcal{T}$ & 10 N\tablenotemark{d} & Thrust\\
$T$ & $0.65$ & End-to-end facility throughput, excluding photometric aperture factor \\
$X$ & $0.7$ & Photometric aperture radius in $\lambda/D$ \\
$\Upsilon$ & $0.69$ & Fraction of Airy pattern contained within photometric aperture \\
$\Omega$ & $b\pi(X\lambda/D)^2$ radians\tablenotemark{\dagger} & Solid angle subtended by photometric aperture \\
$b$ & $1.0$ & Areal broadening of the planet's PSF \\
IWA & 60 mas & Inner working angle \\
OWA & $\infty$ & Outer working angle \\
$\Delta$mag$_{\rm floor}$ & $2.5 - 2.5\log{\zeta_{\rm d}}$\tablenotemark{\dagger} & Systematic noise floor (i.e., dimmest point source detectable at S/N)\\

\hline
$\lambda_{\rm d}$ & $0.55$ $\mu$m & Central wavelength for detection (V band) \\
$\Delta \lambda_{\rm d}$ & $0.22$ $\mu$m & Bandwidth for detection \\
S/N$_{\rm d}$ & $7$ & Signal to noise ratio required for broadband detection of a planet \\
$\zeta_{\rm d}$ & $10^{-10}$ & Raw contrast level in detection region, relative to theoretical Airy pattern peak \\
${\rm CR}_{\rm b,detector,d}$ & $0$ & Detector noise count rate for detection \\
\hline

$\lambda_{\rm c}$ & $1.0$ $\mu$m & Central wavelength for spectral characterization \\
$R_{\rm c}$ & $50$ & Spectral resolving power required for spectral characterization \\
S/N$_{\rm c}$ & $5$ & Signal to noise ratio per spectral resolution element required for spectral characterization \\
$\zeta_{\rm c}$ & $10^{-10}$ & Raw contrast level for spectral characterization, relative to theoretical Airy pattern peak \\
${\rm CR}_{\rm b,detector,c}$ & $0$ & Detector noise count rate for spectral characterization\\
\enddata
\tablenotetext{a}{Optimized value chosen by code given high thrust assumption.}
\tablenotetext{b}{Optimized for a given bandpass, IWA, contrast, and telescope diameter; see Section \ref{ss_design_section}.}
\tablenotetext{c}{See Section \ref{ss_design_section}.}
\tablenotetext{d}{Value chosen to ensure starshade operates far from thrust-limited regime.}
\tablenotetext{\dagger}{Dependent quantity}
\end{deluxetable}

\subsection{Uncertainty in Yield}

We note that all yield estimates are probabilistic quantities.  We don't know ahead of time which stars will actually host the planets, so it's possible that certain instances of nature may be more or less productive than other instances.  Thus, all yield estimates should have an intrinsic uncertainty based on the possible arrangement of planetary systems.  This is uncertainty has been included by some previous studies \cite{savransky2010}.

However, this source of uncertainty is significantly less than other sources, including astrophysical quantities (e.g., $\eta_{\rm Earth}$ and median exozodi level), un-modeled instrument effects (e.g., contrast as a function of stellar diameter), and the as-yet unstudied impact of observational techniques and planet discrimination methods (e.g., characterization by orbit constraint vs characterization by spectroscopy).  Because of this, we choose not to plot the uncertainties associated with planetary arrangement alone.  However, we emphasize that the yield estimates presented in this work will likely change as these parameters and effects are better understood, and that each yield is accompanied by an intrinsic uncertainty.

\subsection{Starshade design\label{ss_design_section}}

We adopted numerically optimized starshades designed to produce the adopted IWA and contrast over the full wavelength band from $0.5$--$1.0$ $\mu$m, while minimizing both the starshade diameter and telescope-starshade separation \cite{vanderbei2007,cady2009}.  To numerically design a starshade, we approximated the electric field near the center of the starshade shadow as the field following an azimuthally-symmetric apodization, and used a linear optimization to produce a radial profile that meets any desired input constraints.  With a sufficient number of petals, this approximation to a starshade with a shaped edge becomes arbitrarily good.
 
While this optimization is usually done with physical parameters---lengths, wavelengths, angular distances---these parameters have some degeneracies. For example, identical starshade shapes may be used at different distances from the telescope if a different IWA and bandpass are assumed.  We rewrote the optimization in terms of four non-dimensional, non-degenerate parameters, enabling efficient coverage of the parameter space of possible optimized starshade designs \cite{cady2011}.  We then interpolated over this grid to estimate the diameter and distance of a starshade defined by a given IWA, contrast, and telescope diameter.

We note that starshade contrast can vary with stellar separation while our DRM code adopts a singular uniform contrast exterior to the IWA.  To simplify the problem, we adopted a singular contrast metric for each starshade design, evaluated by averaging the contrast over a 1 $\lambda/D$ wide annulus centered at 1 $\lambda/D$ exterior to the IWA.

For each starshade, we calculated the dry mass as
\begin{equation}
\label{dry_mass_equation}
	m_{\rm dry} = (1+\mathcal{C}) \times \left(m_{\rm ss} + m_{\rm bus} \right),
\end{equation}		
where $\mathcal{C}=0.3$ is the contingency factor, $m_{\rm ss}$ is the starshade payload, and $m_{\rm bus}$ is the spacecraft bus.  For the starshade payload mass, we used a linear fit to recent starshade designs \cite{exos2015}, such that 
\begin{equation}
\label{starshade_mass_equation}
	m_{\rm ss} = 35\times\left(\frac{D_{\rm ss}}{1\; {\rm m}}\right) - 620\; {\rm kg}.
\end{equation}
For the bus mass we adopted the following formula, based on the analysis of Ref.~\citenum{exos2015}, which scales bus mass with starshade and fuel mass:
\begin{equation}
\label{bus_mass_equation}
	m_{\rm bus} = 615 + 0.09 \left( \frac{m_{\rm ss}}{1\; {\rm kg}} - 560 \right) + 0.03 \left(\frac{m_{\rm tot}}{1\; {\rm kg}} - \frac{m_{\rm dry}}{1\; {\rm kg}} \right)\; {\rm kg}.
\end{equation}
Substituting Equations \ref{starshade_mass_equation} and \ref{bus_mass_equation} into Equation \ref{dry_mass_equation} gives the final dry mass equation we used,
\begin{equation}
	m_{\rm dry} \approx 52 \times \left(\frac{D_{\rm ss}}{1\; {\rm m}}\right) + 0.04 \times \left(\frac{m_{\rm tot}}{1\; {\rm kg}}\right) - 154\; {\rm kg}.
\end{equation}

\section{Methods}
\label{methods_section}

To maximize the yield of a coronagraph-based mission, \starketal\cite{stark2015} optimized the use of the single limiting resource, exposure time.  To maximize the yield of a starshade, we must recognize that starshades have two limiting resources: exposure time and fuel.  For a fixed-length mission, these two resources are not independent; fuel use depends on slew time, which is the fraction of the mission not used for exoplanet science exposures.  For example, one could reduce fuel consumption per target by sacrificing exposure time and awarding it to slew time, allowing one to visit more targets for the same total fuel mass.

To better understand this trade-off, we find it useful to first proceed in Section \ref{time_budgeted_section} with a simplistic expression for fuel use and optimize yield under the incorrect assumption that time and fuel are independent resources.  We refer to this scenario as the ``time-budgeted" optimization scenario.  We will then present an approximate analytic expression for fuel use in Section \ref{fuel_use_expression_section}, relating fuel mass to exposure time, and show that the ``time-budgeted" optimization does not ideally balance fuel use with exposure time.  Finally, in Section \ref{fuel_time_balanced_section} we will detail how to arrive at an ideally optimized starshade yield for a given fuel mass by using our analytic fuel use expression to balance between slew time, exposure time, number of slews, and number of targets. We refer to this latter method as the ``fuel-time balanced" optimization method.

\subsection{Time-budgeted optimization\label{time_budgeted_section}}

\starketal\cite{stark2014} discusses an algorithm to maximize the completeness, $C$, and thus yield, by optimizing the observation plan.  This algorithm consists of the following steps:
\begin{enumerate}
\item Award exposure time $d\tau$ to the $i$th observation with the largest completeness curve slope, $dC_i/d\tau$, where $\tau$ is exposure time.  As a result of this process, all observations will achieve the same final completeness curve slope.
\item Prioritize observations by the benefit-to-cost ratio $C_i/\tau_i$, then down-select to those observations that fit within the total exposure time budget.
\end{enumerate}

The prioritization in the second step assumes that the cost of an observation is simply the exposure time, which is valid for a coronagraph-based mission where time is the limiting resource. Starshades, however, have an additional limiting resource: fuel.  The cost metric for a starshade observation is therefore some combination of fuel and exposure time.  In practice, calculating the fuel associated with an observation using a starshade is non-trivial and depends on many factors, including the propulsion assumed, station keeping during observation, number of slews, distance between each slew, and the time allowed for each slew.  Because our code does not calculate an observation schedule, we do not know the precise distance of any single slew of the starshade, and it is therefore technically impossible to prioritize the observations by the actual fuel usage.

In light of this, we can simplify the problem and approximate the fractional fuel used by a single observation as $1/n_{\rm slews}$, where $n_{\rm slews}$ is the number of starshade slews performed.  Under the assumption that each slew consumes fractional fuel equivalent to $1/n_{\rm slews}$, one choice for the cost metric of the $i$th observation could be $1/n_{\rm slews} + \tau_i/\Sigma\tau$, where $\Sigma\tau$ is the total exposure time budget.  However, this cost fails to capture the slew-limited regime: if $n_{\rm slews}$ is sufficiently small, the mission may not use up all of the exposure time, in which case exposure time should not be considered a cost.  For example, imagine a scenario in which only 10 slews of the starshade are allowed for the baseline mission.  In this case, the baseline mission could completely search the observable portion of 10 HZs to the noise floor in far less than the exposure time allowed.  In the slew-limited regime (i.e., fuel-limited regime), the exposure time of any individual target is irrelevant and the cost metric should be the fuel consumption only.

To capture the transition between the slew- and exposure time-limited regimes, we can adopt a linear combination of fractional fuel usage and fractional exposure time usage, such that the benefit-to-cost ratio for the $i^{\rm th}$ observation is given by
\begin{equation}
	\frac{\rm Benefit}{\rm Cost} = \frac{C_i}{\kappa/n_{\rm slews} + \left(1-\kappa\right) \tau_i / \Sigma\tau},
\end{equation}
where $0<\kappa<1$.  We can allow the code to choose the value of $\kappa$ that maximizes the yield by calculating the yield for 1000 different values of $\kappa$.  In practice, this negligibly increases the run time of the yield estimation code, as the majority of the run time is spent on other optimizations like the first step listed above.  We note that we tried other cost metrics, including raising the fuel and exposure time usage to optimized powers, but found the linear combination above to achieve the highest yields in the transition between the fuel- and exposure time-limited regimes.

By adopting $1/n_{\rm slews}$ as the fractional fuel used per slew, we are setting the fuel consumption of each slew equal to the average fuel per slew, a reasonable approximation for many slews.  A more detailed simulation would take into account the path of the starshade and optimize the fuel usage.  However, keep in mind that our fuel cost metric does not say anything about the absolute amount of fuel used by the mission, a calculation we will address later.

\begin{figure}[H]
\begin{center}
\includegraphics[width=4in]{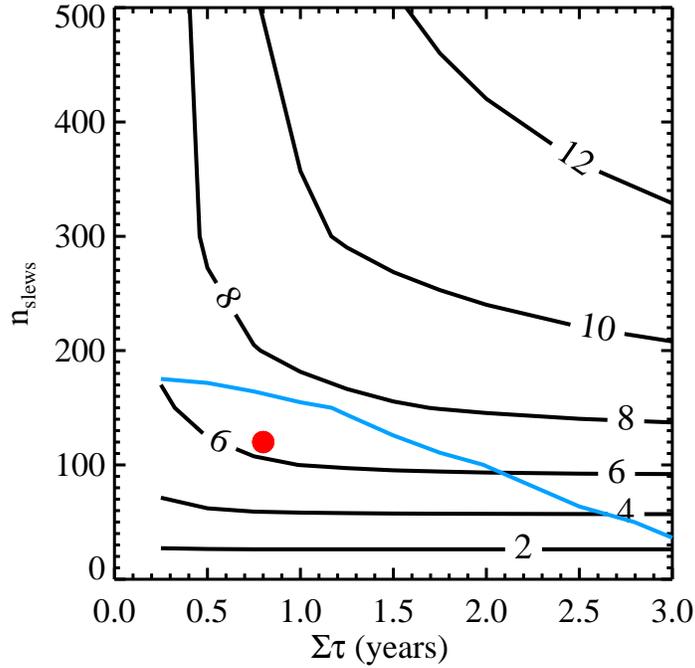}
\caption{ExoEarth candidate yield contours for the baseline mission (black) as functions of total exposure time and number of starshade slews using the non-ideal time-budgeted optimization method.  The red dot marks the location of the baseline mission. A hypothetical contour of constant fuel mass (blue) illustrates that one can trade off exposure time for slews to increase yield; time-budgeted optimization does not necessarily produce the ideal solution. \label{yield_contours_baseline_figure}}
\end{center}
\end{figure}

We implemented the above method and calculated the optimized yield as a function of the two limiting resources $n_{\rm slews}$ and $\Sigma\tau$.  Figure \ref{yield_contours_baseline_figure} shows yield contours for the baseline mission parameters in black.  In this plot, the $y$-axis corresponds to the number of slews possible---the number of slews chosen by the code that maximizes the yield could be smaller.

Each yield contour has horizontal and vertical asymptotes.  The horizontal asymptotes correspond to the slew-limited regime, in which additional photons do not produce higher yields because $\sim n_{\rm slews}$ HZs have been observed as completely as possible.  The vertical asymptotes correspond to the exposure time-limited regime, in which additional slews do not produce higher yields because the mission does not have enough exposure time to use them.  The elbow at the lower-left of each curve represents the transition between the fuel- and exposure time-limited regimes, where a starshade-based mission can efficiently use all of the allowed slews and all of the exposure time simultaneously.

Of course for a given mission, we are not allowed any combination of $n_{\rm slews}$ and $\Sigma \tau$ that we desire---this choice is governed by the available fuel mass and usage.  Figure \ref{yield_contours_baseline_figure} plots a single hypothetical fuel mass contour in blue, roughly based on the equations presented in the following section.  The blue curve illustrates that the available fuel mass should determine the precise combination of $n_{\rm slews}$ and $\Sigma \tau$ that maximize yield, an additional optimization that our time-budgeted method does not perform.

\subsection{Analytic expressions for fuel use\label{fuel_use_expression_section}}

As shown in the previous section, to maximize the yield of a starshade mission we need to optimally balance exposure time and the number of slews for a given mission.  Because the total slew time is the remainder of the mission not used for exposures, this balancing act requires relating the number of slews to the total slew time, i.e., an expression for fuel use.  Here we provide a method to calculate fuel use based on simple scaling relationships.

Basic rocket physics tells us
\begin{equation}
\label{rocket_equation}
	\ln{\frac{m + dm}{m}} = \frac{\Delta v}{v_{\rm ex}},
\end{equation}
where $m$ is initial mass, $dm$ is the change in mass (fuel expended), $\Delta v$ is the change in velocity of the starshade, and $v_{\rm ex}$ is the exhaust velocity of the propellant.  Due to concerns about exhaust from an ion propulsion system interfering with observations, recent starshade designs have assumed separate propulsion systems for slewing and stationkeeping, adopting an efficient ion propulsion for the former and chemical propulsion for the latter \cite{savransky2010,exos2015}.  Separating these two factors we can rewrite the above expression in terms of the specific impulse of each propulsion system as
\begin{equation}
\label{rocket_equation2}
	\ln{\frac{m + dm}{m}} = \frac{\Delta v_{\rm slew}}{g\, I_{\rm slew}} + \frac{\Delta v_{\rm sk}}{g\, I_{\rm sk}},
\end{equation}
where $g=9.8$ m s$^{-2}$.  Below we present approximate analytic expressions for fuel used by slewing and stationkeeping for each observation.  We include only the terms that are commonly included in the literature, neglecting radiation pressure on the starshade and the contribution of the orbital dynamics to the slew time.  We then explain how to use these expressions to estimate mission-long fuel use.

\subsubsection{Stationkeeping}

\begin{figure}[H]
\centering
\includegraphics[width=6in]{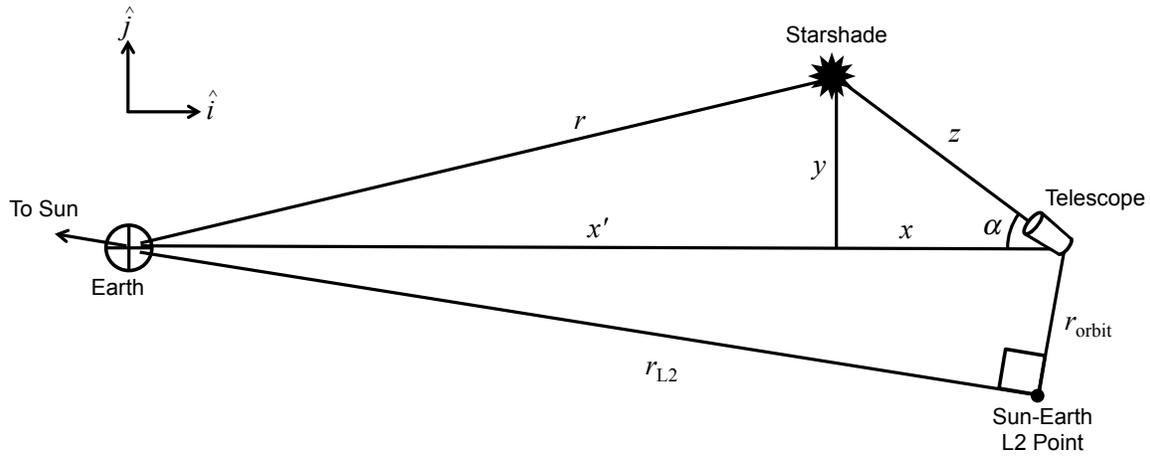}
\caption{Diagram of telescope and starshade orientation with respect to Earth observing at an angle $\alpha$.  Because $r_{\rm orbit} \ll 1$ AU, $\alpha$ is approximately the solar elongation.  \label{ss_orbit_schematic}}
\end{figure}

Figure \ref{ss_orbit_schematic} shows the approximate geometry of the starshade and telescope orbiting about L2 at a distance $r_{\rm orbit}$ (exaggerated for illustration), as seen from the North ecliptic pole at a time when they are located in the ecliptic plane.  For a more precise illustration showing the starshade and telescope orbits, see Figure 2 in Ref.~\citenum{kolemenkasdin2007}.  Given that $r_{\rm orbit} \ll 1$ AU, solar elongation $\approx\alpha$.

The starshade and telescope must be kept aligned to within roughly a meter or so in the lateral direction to ensure that the telescope stays within the starshade's shadow \cite{exos2015}.  While the tolerance is much looser in the axial direction ($\hat{z}$), the distance offset must be controlled at some point, either as part of stationkeeping or during the slew of the telescope.  For simplicity, we decide to include the axial disturbances in the stationkeeping fuel use budget.

In a non-inertial rotating reference frame, such as that rotating with L2 about the Sun-Earth barycenter, we can express the acceleration felt by any object as
\begin{equation}
\label{rotrefacc}
	\vec{a^{\prime}} = \vec{a} - \vec{\Omega} \times \left(\vec{\Omega} \times \vec{r}\right) - 2 \vec{\Omega} \times \vec{v^{\prime}},
\end{equation}
where the first term $\vec{a}$ is the acceleration felt in the inertial frame due to real forces, the second term is the centrifugal pseudo-force term, the third term is the Coriolis pseudo-force term, $\vec{\Omega}$ is the angular rotation rate vector of the rotating frame, $\vec{r}$ is the position of the object in the inertial frame, and $\vec{v^{\prime}}$ is the velocity of the object in the rotating frame.

For the moment, we will ignore the fact that the telescope and starshade are orbiting about L2 and adopt the orientation illustrated by Figure \ref{ss_orbit_schematic}.  For both the starshade and the telescope, which are assumed to be orbiting the Sun at the same rate as Earth, the Sun's gravitational pull should approximately cancel out with each objects' centrifugal term.  Thus, we are left with terms corresponding to Earth's gravitational pull and the Coriolis force.  We start with the Coriolis force.

Both the telescope and starshade are moving on an orbit about L2, such that they both have velocities in the reference frame rotating with Earth about the Sun and therefore both have associated Coriolis terms.  Because we are ultimately interested in the \emph{difference} in acceleration between the two spacecraft, and to first order their velocities are the same, the majority of the Coriolis effect cancels out.  However, the starshade and telescope will have a non-negligible difference in velocity to maintain alignment with a given star.  During an observation, the telescope must change its pointing and the starshade must move relative to the telescope to track the motion of the stars.  Given the orientation shown in Figure \ref{ss_orbit_schematic}, this means the starshade must have a velocity relative to the telescope in the rotating reference frame which is roughly directed along $\hat{i}$.  Thus, we can roughly express the Coriolis effect on the starshade as
\begin{equation}
	\vec{a}_{\rm SS,coriolis} \approx 2 \Omega v_{\rm SS} \hat{k},
\end{equation}
where $v_{\rm SS}$ is the speed of the starshade relative to the telescope in the $\hat{i}$ direction and $\hat{k}$ is directed out of the page.  To track the sky when both spacecraft are in the ecliptic, $v_{\rm SS} \approx \Omega z$, such that
\begin{equation}
	\vec{a}_{\rm SS,coriolis} \approx 2 \Omega^2 z \hat{k}.
\end{equation}
Given that $\Omega = 2\pi$ yr$^{-1}$, or $\approx 2\times10^{-7}$ s$^{-1}$, the difference in acceleration in the frame rotating with Earth between the starshade and telescope due to the Coriolis effect is $\approx 1.6\times10^{-7} (z/1\,{\rm Mm})$ m s$^{-2}$ for targets in the ecliptic.  When looking at targets at the ecliptic poles, $\vec{\Omega}$ and $\vec{v_{\rm SS}}$ are antiparallel, such that there is no difference in the Coriolis terms.  Assuming targets are uniformly distributed over the sky, such that the median target is at an absolute declination of 30$^{\circ}$, the median difference in acceleration due to the Coriolis effect should be roughly a factor of $\cos{30^{\circ}}$ smaller, such that $a_{\rm SS,coriolis} \approx 1.4\times10^{-7} (z/1\,{\rm Mm})$ m s$^{-2}$.

Now we address the first two terms of Equation \ref{rotrefacc}.  Because we are interested in the difference in the forces acting on the starshade and telescope, and because the centrifugal terms cancel with the Sun's gravitational force, the difference in the remaining forces can be well-approximated by the difference in Earth's gravitational influence on the two spacecraft.  These terms can therefore be approximated as
\begin{equation}
	\vec{a}_{\rm T} = -\frac{GM_{\Earth}}{r'\,^2} \hat{i},
\end{equation}
and
\begin{equation}
	\vec{a}_{\rm SS} = -\frac{GM_{\Earth}}{r^2} \hat{r}.
\end{equation}
Taking the difference of these two vectors expressed in terms of $\hat{i}$ and $\hat{j}$, and substituting $\mathcal{Z} = z/r'$, where $r' = \sqrt{r_{\rm L2}^2 + r_{\rm orbit}^2}$ gives
\begin{equation}
	\Delta \vec{a} \approx \frac{GM_{\Earth}}{\left(r_{\rm L2}^2 + r_{\rm orbit}^2\right)} \left( 1 - \frac{1 - \mathcal{Z}\cos{\alpha}}{\left(1-2\mathcal{Z}\cos{\alpha}\right)^{3/2}}\,\hat{i} - \frac{\mathcal{Z}\sin{\alpha}}{\left(1-2\mathcal{Z}\cos{\alpha}\right)^{3/2}}\,\hat{j} \right).
\end{equation}
Expanding to first order in $\mathcal{Z}$ and taking the magnitude gives
\begin{equation}
	\Delta a \approx \frac{GM_{\Earth}\, z}{\left(r_{\rm L2}^2 + r_{\rm orbit}^2\right)^{3/2}} \left( 1+3\cos^2{\alpha}\right)^{1/2}.
\end{equation}
Assuming an average solar elongation of $\alpha = 60^{\circ}$ and $r_{\rm orbit} = 0.5 r_{\rm L2}$, we find $\Delta a \approx 1.1\times10^{-7} \left( z / 1\; \rm{Mm} \right)$ m s$^{-2}$ when the starshade and telescope are located in the ecliptic plane.

The difference in direction between the Coriolis and gravitational acceleration terms changes with time and depends on the orbit about L2 and the coordinates of the target star.  Thus, we cannot simply add these two terms in quadrature; improved estimates of the overall relative accelerations require detailed modeling of the orbital dynamics, which is beyond the scope of this paper.  However, the above simple estimates illustrate that the dominant terms controlling the stationkeeping fuel budget are both proportional to starshade-telescope separation, $z$, and of order $\sim1\times10^{-7}(z/1\,{\rm Mm})$ m s$^{-2}$.  This is in agreement with previous studies that have performed such detailed modeling.  For example, \savetal\cite{savransky2010} adopted $\Delta a = 1.4\times10^{-7} \left( z / 1\; \rm{Mm} \right)$ m s$^{-2}$ using the orbit-averaged results of Ref.~\citenum{kolemenkasdin2007}.  For this study, we adopt the same value of $\Delta a$, and express station keeping fuel use for the $i^{\rm th}$ observation as
\begin{equation}
	\frac{\Delta v_{i,{\rm sk}}}{g\, I_{\rm sk}} \approx  \frac{\Delta a\, \tau_i}{g\, I_{\rm sk}\, \epsilon_{\rm sk}},
\end{equation}
where $\epsilon_{\rm sk}$ is the efficiency of station keeping fuel use, typically $\sim 0.8$ due to the cant of the thrusters.

In the above derivation, we neglected radiation pressure.  At the Sun-Earth L2 point, the absolute acceleration due to radiation pressure on a 100 m diameter starshade at a solar elongation of $60^{\circ}$, with a mass of 5000 kg, is $a_{\rm RP} \approx 5\times10^{-6}$ m s$^{-2}$.  This is on par with the acceleration difference due to Earth's gravity at a separation of $z = 100$ Mm.  However, the \emph{relative} radiation pressure force will be offset by the pressure on the telescope, which we assume can be modified to make the difference in acceleration due to radiation pressure negligible.

We also neglected the fuel demands of a spinning starshade.  If the starshade must spin for dynamical stability or to mitigate the effects of petal defects, additional fuel will be required to either spin up/down the starshade or change the starshade angular momentum vector before and after slewing.

\subsubsection{Slewing}

\begin{figure}[H]
\centering
\includegraphics[width=6in,trim={1.2cm 6.8cm 8.6cm 7.7cm},clip]{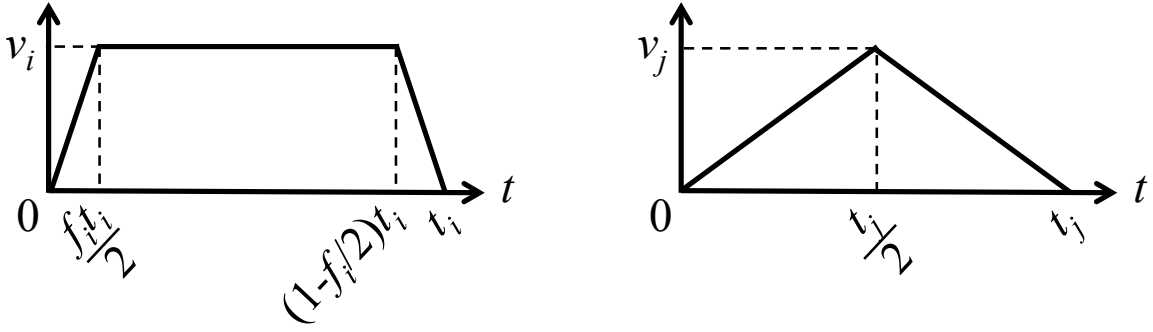}
\caption{Example starshade slewing velocity profiles of an efficient slew (left) and an inefficient, thrust-limited slew (right).  \label{velocity_profile_illustration}}
\end{figure}

While the starshade and telescope are orbiting near L2, the starshade must move relative to the telescope to point at a new target.  Figure \ref{velocity_profile_illustration} illustrates two potential slews for a starshade by plotting velocity as a function of time. We have assumed that the mass of the starshade changes negligibly during a given slew, such that the initial and final accelerations are equal and the velocity profiles are symmetric. On the left-hand plot, a small fraction $f_i$ of the total slew time of the $i^{\rm th}$ slew is spent accelerating/decelerating, such that the majority of the slew is spent coasting; this velocity profile represents efficient use of starshade fuel. On the right, $f_j=1$ and the starshade accelerates/decelerates for the entire duration of the slew; this velocity profile represents inefficient thrust-limited use of starshade fuel. 

Assuming the starshade is not operating in the thrust-limited regime, such that $0 < f_i < 1$, we can express the distance traversed during the $i^{\rm th}$ slew as
\begin{equation}
\label{first_slew_distance_equation}
	s_i = \frac{1}{2}a_i \left(\frac{t_i f_i}{2}\right)^2 + v_i t_i \left( 1 - f_i \right) +  \frac{1}{2}a_i \left(\frac{t_i f_i}{2}\right)^2.
\end{equation}
Given that $v_i = a_i t_i f_i / 2$, and $\langle v \rangle_i = s_i / t_i$, we can rewrite this as
\begin{equation}
\label{slew_efficiency_equation}
	\langle v \rangle_i = v_i \left(1-\frac{f_i}{2}\right),
\end{equation}
and note that the slew efficiency $\epsilon_{i,{\rm slew}} = (1-f_i/2)$, where $0<f_i<1$.  

We can also rewrite Equation \ref{first_slew_distance_equation} as
\begin{equation}
\label{second_slew_distance_equation}
	s_i = \frac{1}{2}a_i t_i^2 \left(f_i - \frac{f_i^2}{2}\right).
\end{equation}
Substituting $a_i = \mathcal{T} / m_i$ for the starshade acceleration, where $\mathcal{T}$ is thrust and $m_i$ is the starshade mass during the $i^{\rm th}$ slew,
\begin{equation}
\label{slew_distance_equation}
	s_i = \frac{1}{2}\frac{\mathcal{T}}{m_i} t_i^2 \left(f_i - \frac{f_i^2}{2}\right).
\end{equation}

We make a common assumption used by other studies to date: the slew efficiency is constant for all slews, such that $f_i = f$ and $\epsilon_{i,{\rm slew}} = \epsilon_{\rm slew}$.  Given Equation \ref{slew_distance_equation}, this means that the slew time of the $i^{\rm th}$ slew is given by
\begin{equation}
\label{slew_time_equation}
	t_i = \sqrt{\frac{2 m_i s_i}{\mathcal{T} \left(f - f^2/2\right)}}.
\end{equation}

\begin{figure}[H]
\begin{center}
\includegraphics[width=6in]{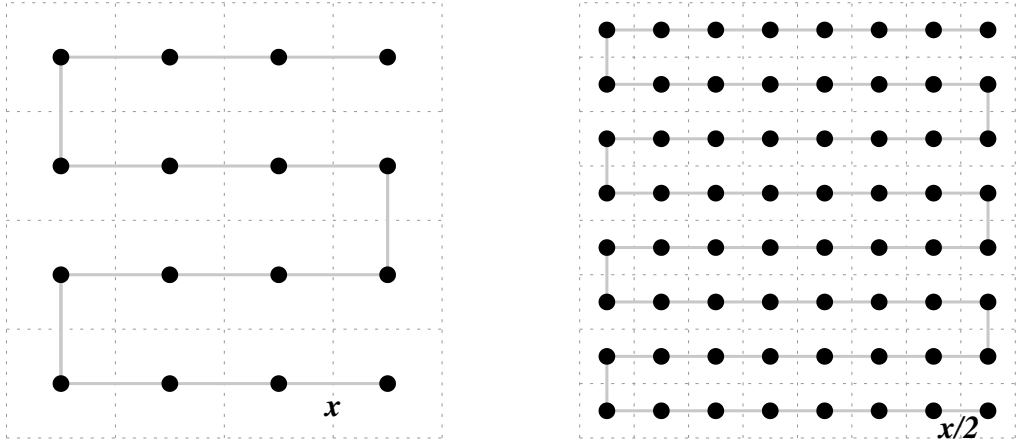}
\caption{Diagrams illustrating how the maximally efficient path (gray line) changes as the number of uniformly distributed targets (dots) increases.  The diagram on the right has 4 times as many targets, but the slew distance is only half that of the diagram on the left.  \label{distance_diagram}}
\end{center}
\end{figure}

In reality each $s_i$ will be different and will depend on optimization of the starshade's path.  However, we are interested in a scaling relationship that is approximately valid for the mission as a whole.  Thus, we approximate $s_i \approx s$, where $s$ is the typical slew distance.  We approximate the typical slew distance by assuming that $n_{\rm targets}$ are distributed uniformly over the sky.  Under this assumption, we can divide the sky up into $n_{\rm targets}$ patches, similar to what is shown by the left diagram in Figure \ref{distance_diagram}.  The gray line in this diagram traces a maximally efficient path through these 16 targets, where each slew requires a slew distance of $x$.  The diagram on the right shows that if we quadruple the number of points to an 8$\times$8 grid, a maximally efficient path requires a slew distance of $x/2$.  Thus the slew distance is proportional to the inverse square root of the number of targets. 

Assuming the starshade travels along the chord between targets distributed uniformly over the surface of a sphere of radius $z$, where $z$ is the separation between the starshade and telescope,
\begin{equation}
\label{typical_slew_distance_equation}
	s = 2z\sin{\sqrt{\frac{\pi}{n_{\rm targets}}}}.
\end{equation}

Equation \ref{slew_efficiency_equation} allows us to express the $\Delta v$ required for the $i^{\rm th}$ slew ($=2v_i$) as
\begin{equation}
\label{dv_slew_equation}
	\frac{\Delta v_{i,{\rm slew}}}{g\, I_{\rm slew}} = \frac{2 s_i}{g I_{\rm slew} \epsilon_{\rm slew} t_i}.
\end{equation}
Equations \ref{slew_time_equation} and \ref{dv_slew_equation}, with $s_i = s$ given by Equation \ref{typical_slew_distance_equation}, constitute our final slew fuel use expressions.

\subsubsection{Caveats\label{caveats_section}}

In the above derivations, a number of assumptions and approximations were made.  First, Equation \ref{typical_slew_distance_equation} assumed that targets were uniformly distributed over the sky and that a maximally efficient path would be chosen.  In reality, the starshade may in fact back track or otherwise deviate from an optimal path due to pointing constraints and limited knowledge of the ideal path.  This may lead one to believe that we will underestimate the amount of fuel required.  However, we also note that detailed starshade DRMs, like those of \savetal\cite{savransky2010}, \emph{choose} targets in part based upon their slew distance.  As a result, targets may not be uniformly distributed over the sky and the typical slew distance may in fact be smaller than we estimate.  Additionally, our slew estimates do not calculate the detailed quasi-halo orbital dynamics of the telescope and starshade \cite{kolemenkasdin2007}.

We also remind the reader that our calculations do not simulate the execution of a mission like those of \savetal\cite{savransky2010}. We do not model the detailed pointing of the starshade/telescope, nor do we ``schedule" each observation given a defined field of regard. Instead, we simply adopt the expected median solar elongation of $60^{\circ}$ for each target and assume all observations are schedulabe.  To prevent yield gains due to unrealistic timing optimizations, we choose not to optimize the time between revisits.

\subsubsection{Fuel use calculation method\label{final_fuel_use_expressions_section}}

To calculate the fuel mass for a set of ($n_{\rm slews}$, $n_{\rm targets}$, $\Sigma\tau$, and $\Sigma t_{\rm slew}$), we employ the following procedure, starting with the dry mass $m_{\rm dry} = m_0$:
\begin{enumerate}
	\item Calculate the starshade mass after the $i^{\rm th}$ observation's station keeping as
	\begin{equation}
		m_i' = m_{i} \exp{\left(\frac{\Delta a_{\rm transverse}\, \tau_i}{g\, I_{\rm sk}\, \epsilon_{\rm sk}}\right)},
	\end{equation}
	\item Calculate the slew distance for the $i^{\rm th}$ slew $s_i = s$ given by Equation \ref{typical_slew_distance_equation}.
	\item Calculate the slew time for the $i^{\rm th}$ slew $t_i$ using Equation \ref{slew_time_equation} with $m_i'$ above for $m_i$.
	\item Calculate the starshade mass after the $i^{\rm th}$ observation's slew as
	\begin{equation}
		m_{i+1} = m_i' \exp{\left(\frac{2\, s}{g\, I_{\rm slew}\, \epsilon_{\rm slew}\, t_i}\right)},
	\end{equation}
\end{enumerate}
The above procedure is calculated until $i=n_{\rm slews}$, at which point the total mass is known, and the fuel mass can be determined by subtracting off the dry mass.

We compared our fuel mass calculations to 100 individual simulations of a single starshade design by \savetal\cite{savransky2010}.  When adopting the starshade parameters and lateral disturbance acceleration used by \savetal\cite{savransky2010}, our estimated fuel mass was $4\%$ larger than that calculated by \savetal\cite{savransky2010} on average, with a standard deviation of 4\%, and produced an average slew time that fit with the 5 year budget adopted by \savetal\cite{savransky2010}.  Given the exponential nature of the fuel mass expression, we consider this to be good agreement.

\subsection{Fuel-time balanced optimization\label{fuel_time_balanced_section}}

Now that we have illustrated the importance of balancing exposure time with slew time to maximize yield, we discuss an alternative, preferred method for yield maximization that automates this balancing process.  In Section \ref{time_budgeted_section} we modified the benefit-to-cost ratio to take into account fractional fuel use, treating each slew equally.  While it may seem reasonable to consider fuel \emph{mass} as our limiting fuel resource, this would require knowing the order of observations before planning them, as the fuel mass expended for the $i^{\rm th}$ observation depends on the mass of the starshade during the $i^{\rm th}$ observation.  Additionally, because of the exponential decay of starshade mass, it is actually the mass ratio of each observation that matters.  Therefore we consider the $\Delta v_i / v_{\rm ex}$ (the righthand side of Equation \ref{rocket_equation}) our limiting resource.

Under this assumption, and demanding that we ignore the order of observations, only the station keeping term, which is linearly proportional to $\tau_i$, is unique to each observation.  As a result, our method of distributing exposure time $d\tau$ to each observation such that all observations have equal slopes $dC_i/d\tau_i$, detailed in Step 1 of Section \ref{time_budgeted_section}, is still valid and simultaneously optimizes both the fuel and exposure time resource distributions under the assumption that all slews are equal.

However, we still must optimize the balance between total exposure time and total slew time.  To do so, we implement the following procedure, a modification to the ``equal-slope" method of Ref.~\citenum{hunyadi2007}:
\begin{enumerate}
	\item Given a set of mission parameters, calculate the completeness curves of every observation, $dC_i/d\tau$.
	\item Guess the optimal slope of all completeness curves, $(dC/d\tau)_{\rm o}$.
	\item \label{exposure_time_step} Find the exposure time $\tau_i$ for each observation at which $dC_i/d\tau = (dC/d\tau)_{\rm o}$.
	\item \label{effslew_step} Assume $\epsilon_{\rm slew} = 0.5$.
	\item \label{ntarg_nslew_step} Assume a valid combination of $n_{\rm targets}$ and $n_{\rm slews}$, such that $n_{\rm targets} \le n_{\rm slews}$, starting with values smaller than the optimal values.
	\item \label{prioritization_step} Prioritize observations by the benefit to cost ratio at the optimized exposure time, $C_i(\tau_i)/X_i$.
	\item \label{yield_step} Select all top priority observations that fulfill the $n_{\rm targets}$ and $n_{\rm slews}$ constraints. Calculate the total observation plus slew time, as well as the total starshade mass.  If both are within the limits of your mission parameters, determine yield $\eta_{\rm planet}\Sigma C_i$ and store this valid solution.
	\item \label{kappa_loop_step} Repeat steps \ref{prioritization_step}--\ref{yield_step} while varying $\kappa$ over a fine grid of values spanning $0\le \kappa \le 1$.
	\item \label{ntarg_loop_step} Repeat steps \ref{ntarg_nslew_step}--\ref{kappa_loop_step} while varying $n_{\rm targets}$ and $n_{\rm slews}$.
	\item \label{effslew_loop_step} Repeat steps \ref{effslew_step}--\ref{ntarg_loop_step}, covering all possible values of $\epsilon_{\rm slew}$.
	\item Repeat steps \ref{exposure_time_step}--\ref{effslew_loop_step} to optimize the completeness curve slope.
	\item Select the combination that produces the highest yield.
\end{enumerate}

\begin{figure}[H]
\centering
\includegraphics[width=6.5in]{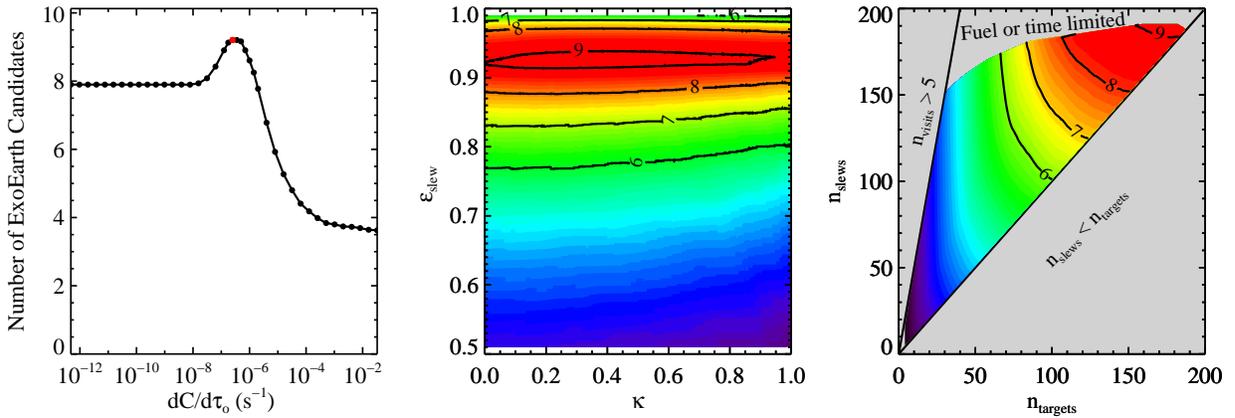}
\caption{The 5 dimensional space over which our code maximizes yield for the baseline mission. Left: yield as a function of completeness curve slope.  The red dot marks the global yield maximum after scanning the remaining 4 parameters. Middle: yield contours for two of the remaining parameters.  All points in this plot correspond to the maximum achieved after scanning the remaining two parameters and represent the expanded dimensionality of the red point in the left plot. Right: yield contours of the remaining two parameters, representing the expanded dimensionality of a single point in the middle plot.  \label{optimization_figure}}
\end{figure}

The above algorithm essentially describes nested for loops that are designed to scan the 5 dimensional parameter space controlling yield. Figure \ref{optimization_figure} illustrates this multi-parameter space for our baseline mission parameters.  The plot on the left shows the yield as a function of $(dC/d\tau)_{\rm o}$.  At each point in this plot, the code scanned the 4 dimensional $(\kappa,\epsilon_{\rm slew},n_{\rm targets},n_{\rm slews})$ subspace and returned the maximum from that subspace.  The red dot represents the final yield, maximized over all 5 parameters.

Because we cannot easily illustrate the 4 dimensional subspace, we must show projections of the yield. The middle panel shows a contour projection of the yield in the $(\kappa,\epsilon_{\rm slew})$ subspace, where each yield value represents the maximum from the remaining two parameters, $n_{\rm targets}$ and $n_{\rm slews}$.  The middle panel corresponds to the complete parameter space scanned for just the red dot in the left plot.

Finally, the right contour plot shows the yield of the $(n_{\rm targets}, n_{\rm slews})$ subspace.  All yields in this contour plot correspond to the single point at the maximum of the middle panel.  The region to the bottom right, with $n_{\rm slews} < n_{\rm targets}$, is excluded, as you must have at least as many observations as targets. The region to the top-left is excluded by our 5 visit limit.  The region at the top is excluded by fuel limitations.  All yield surfaces are well behaved and an obvious global maximum exists.  Our code indeed scans all relevant parameter space and achieves yield maximization.

The above optimization algorithm is written primarily for clarity.  If one implemented this algorithm explicitly as written, it would require 4 nested loops to cover all parameter space with a sorting routine performed on the innermost loop (for the prioritization metric), producing prohibitively long run times.  In practice, the sorting/prioritization loop can be moved to just interior to the slope optimization loop as long as $\kappa$ is finely sampled.  We find that our choice of 1000 values for $\kappa$ is adequate for all tests performed.

Adding status checks for intelligent loop-breaking and improving the initial value guesses can significantly reduce the run time of the above algorithm.  Doing so, we are able to run the above algorithm with our baseline mission parameters in 3 minutes on a single $2.9$ GHz CPU.  These starshade yield calculations are on par with the coronagraph calculations performed by \starketal\cite{stark2015}; the increase in run time due to the above optimization is mostly offset by the limited number of revisits to each star and lack of revisit optimization that was employed by \starketal\cite{stark2015}.

\section{Results}
\label{results_section}

We have implemented the algorithm above in our DRM code to efficiently calculate optimized yields for starshade missions using the baseline parameters and input target list described in Section \ref{caveats}.  Here we compare our results with previous work and calculate the yield for starshade missions as a function of different mission parameters.

\subsection{Comparison with previous work}
\label{savransky_comparison_section}

To check our calculations, we compared our results with those of \savetal\cite{savransky2010}.  We chose to compare our yield to that calculated by \savetal\cite{savransky2010} for the case of $\eta_{\Earth} = 0.5$.  Thus, for this section only, we will deviate from our baseline assumption of $\eta_{\Earth} = 0.1$.

\savetal\cite{savransky2010} calculated the yield of unique Earth-twin detections using an external occulter combined with a 4 m telescope.  \savetal\cite{savransky2010} simulated the operation of a starshade mission from beginning to end, determining a flight path consistent with targets' visibility windows and realistically calculating fuel use for slews and stationkeeping.  \savetal\cite{savransky2010} randomly distributed planets around each star, such that the yield of a given simulation depended on the random arrangement of planets. To estimate the expected (mean) yield as well as the yield uncertainty, \savetal\cite{savransky2010} performed 100 such simulations.  Our yield simulations explicitly calculate the expected mean yield---therefore, we will compare our yield estimates with the average, or expected yield value, reported by \savetal\cite{savransky2010}.

We modified our code to adopt the same assumptions as \savetal\cite{savransky2010}.  We modified our baseline assumptions of an Earth-twin to that used by \savetal\cite{savransky2010}, as well as the HZ definition, distribution of orbits within the HZ, and exozodi definition.  We used a 4 m telescope aperture and adopted identical starshade performance parameters: a $51.2$ m diameter starshade operating at $70.4$ Mm to provide a 59 mas IWA (evaluated at the 50\% throughput point), $10^{-10}$ contrast, throughput of $0.52$, and 20\% bandpass.  We assumed an identical propulsion system ($I_{\rm slew} = 4160$ s, $_{\rm sk} = 220$ s, $\epsilon_{\rm sk} = 0.75$, $\mathcal{T} = 0.45$ N), identical dry mass ($m_{\rm dry} = 4210$ kg), and identical initial mass (6000 kg).  We adopted identical spectral characterization requirements ($R_{\rm c}=70$, S/N$_{\rm c}=11$), PSF scale (factor of $1.5$ greater in solid angle than Airy pattern) and sampling (27 pixels inside of FWHM), detector noise parameters (RN$=3$ pix$^{-1}$ read$^{-1}$, dark current $=0.001$ pix$^{-1}$ s$^{-1}$, $t_{\rm read}=1000$ s), and approximately the same detection S/N $=4$.  

From the top-right panel of Figure 9 in \savetal\cite{savransky2010}, for $\eta_{\Earth} = 0.5$, \savetal\cite{savransky2010} estimated $\approx16 \pm 3$ exoEarth candidates.  This data point corresponds to an average of 118 observations, 89 unique stars, and an average total exposure time of $\sim$0.8 years.  When adopting identical mission parameters and letting our code optimize the observation plan, we calculated a yield of $\approx22$ exoEarth candidates, roughly 35\% higher than the expected yield of \savetal\cite{savransky2010}.  Our optimized observation plan includes 113 observations and 85 unique stars, \emph{fewer} total observations than \savetal\cite{savransky2010}, and $\sim0.8$ years of total exposure time.

To determine the source of this yield discrepancy, we compared the observations calculated by our code to that of \savetal\cite{savransky2010}.  We determined that the observation plans of \savetal\cite{savransky2010} visited a relatively small number of stars many times, as can be seen by the green and red points in Figure 3 of \savetal\cite{savransky2010}.  Roughly 20 total slews of the starshade were devoted to third visits or greater, which contribute little to the total yield.  In contrast, our code chose not to visit any star more than twice.  If one were to take all of the observations spent on visits beyond the second visit, and devote those to new stars, the yield estimate of \savetal\cite{savransky2010} would have increased by $\sim20\%$.  We therefore conclude that the majority of our increase in yield is due to a difference in how revisits are distributed.

The remaining discrepancy in yield can therefore be attributed to time optimization or target selection.  To address the first, we turned off exposure time optimization in our code and forced our code to always observe to the noise floor.  The resulting yield decreased to $21$ exoEarth candidates, implying that exposure time optimization can only explain a 7\% change in yield.

A majority of the yield discrepancy has been explained by differences in revisit selection and exposure time optimization.  The remaining $\sim10\%$ in yield discrepancy is likely a result of target selection differences.  To check, we recalculated our yield while limiting the code to the exact set of stars chosen by \savetal\cite{savransky2010}.  Our yield was reduced to $19$ exoEarth candidates when adhering to the target stars chosen by \savetal\cite{savransky2010}, suggesting that target selection differences can explain the remaining yield differences.

\begin{figure}[H]
\centering
\includegraphics[width=6.5in]{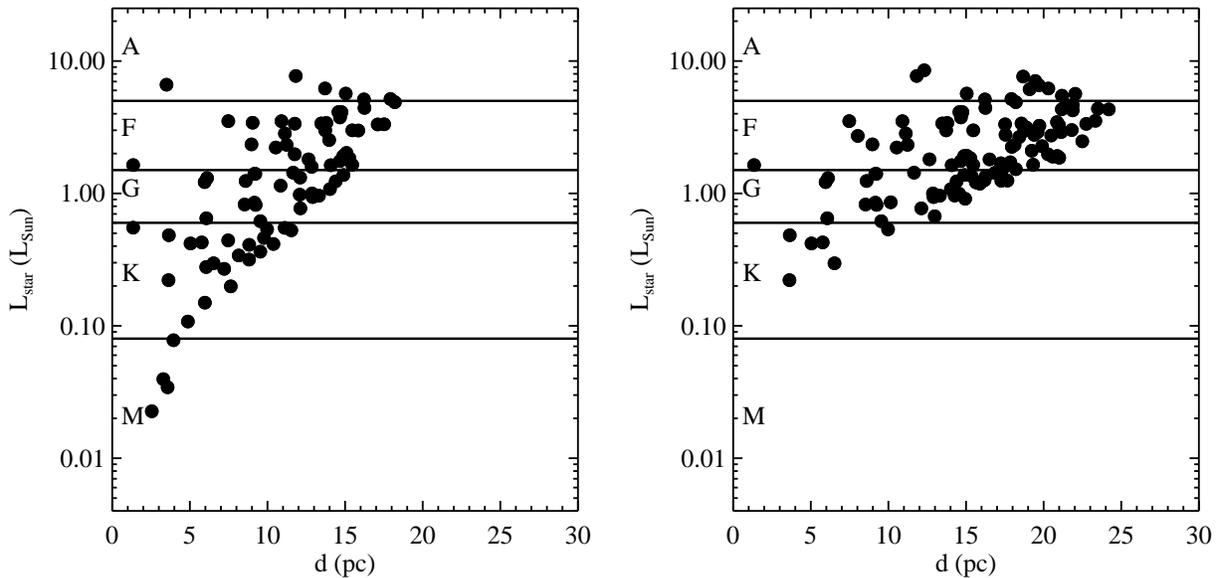}
\caption{Target list comparison with \savetal\cite{savransky2010}.  Left: Targets selected by the DRM described in this paper. Right: Targets selected at least once in 100 DRM simulations by \savetal\cite{savransky2010}.  \label{target_selection_comparison}}
\end{figure}

To determine the source of the target selection differences, we plot the targets selected by our code as well as those selected by all 100 simulations of \savetal\cite{savransky2010} in Figure \ref{target_selection_comparison}.  \savetal\cite{savransky2010}  explicitly removed late K and M stars due to the short orbital periods of HZ planets around those stars.  However, we recalculated our yield while banning stars with $L_{\star} < 0.2$ and found that our yield decreased by a negligible 3\%; low mass stars are not the source of the discrepancy.

Some of the targets selected by \savetal\cite{savransky2010} were identified as giant stars and close-separation binaries by our code.  However, if we were to include these stars in our target list, our yield would only increase.  Giant stars are therefore not the source of the discrepancy.

Thus, the difference in target selection is primarily a small number of nearby F, G, and K stars that have very short exposure times and were not selected by \savetal\cite{savransky2010}, e.g., $\alpha$ Cen B.  It may be surprising that a small handful of stars can change the yield by $\sim$10\%.  However, in this case, $\eta_{\rm Earth}$ was assumed to be 50\%, such that only $\sim4$ nearby stars can add $\sim2$ exoEarth candidates to a yield of $\sim20$.

Of course it's important to note that our code does not explicitly calculate whether a given set of observations is schedulable.  It's therefore possible that some of our yield gain is due to observations that are simply unschedulable.  However, given that our selected targets are more-or-less uniformly distributed over the sky, and that ecliptic latitudes greater than $\sim45^{\circ}$ or less than $-45^{\circ}$ are continuously observable, we find this unlikely.  

\subsection{Yields for starshade missions\label{yields_section}}

We estimated the exoEarth candidate yield for single starshade missions for three different launch mass limits: 3600 kg, 9800 kg, and 25000 kg, corresponding to the Sun-Earth L2 payload mass limits for Falcon 9, Delta IV Heavy, and SLS Block 1 rockets, respectively \cite{rioux2015}.  We do not budget for the telescope mass and allow the starshade to consume the full launch mass limit.  We calculated the yield as a function of telescope aperture size, IWA, and contrast.  We investigated telescope diameters ranging from 2 -- 10 m in steps of 1 m, IWAs ranging from 30 -- 90 mas in steps of 10 mas, and logarithmically-spaced contrasts ranging from $10^{-11}$--$10^{-9}$ for each mass launch limit.  

For each combination of $D$, IWA, and $\zeta$, we interpolated our grid of optimized starshade diameters to determine $D_{\rm ss}$ and $z$.  All other starshade and mission parameters were kept equal to the baseline assumptions listed in Table \ref{baseline_params_table}.  We note that we did not vary the bandpass of the starshade ($0.5$--$1.0$ $\mu$m), which was motivated scientifically.  However, by changing the separation distance of a fixed size starshade, one can reduce the bandpass to improve the IWA, potentially increasing the yield by a modest amount, albeit at a different set of wavelengths \cite{seager2015}.

For every calculation, we adopted our large baseline thrust of 10 N.  This unrealistically large thrust ensured that we were working far from the thrust-limited regime and our yield estimates may therefore be optimistic.  We address this assumption in more detail below.

\begin{figure}[H]
\centering
\includegraphics[width=6.5in]{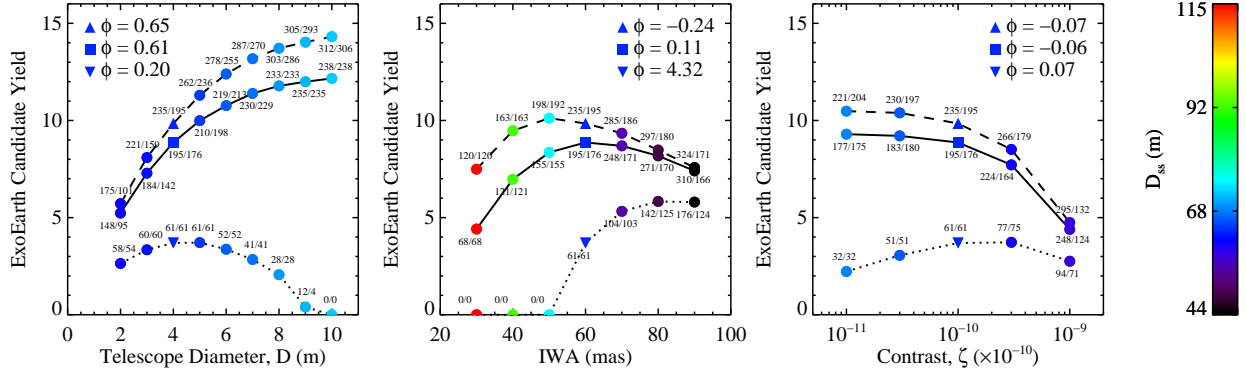}
\caption{ExoEarth candidate yield as a function of telescope and instrument parameters for the baseline mission assuming launch mass limits of 3600 (dotted line), 9800 (solid line), and $25000$ kg (dashed line).  Color indicates starshade diameter and $n_{\rm slews}$/$n_{\rm targets}$ is listed next to each data point. The sensitivity of yield, $\phi$, is given at the baseline parameter values of each curve. \label{yield_vs_instrument_params}}
\end{figure}

The left panel of Figure \ref{yield_vs_instrument_params} shows the exoEarth candidate yield as a function of telescope aperture size, where the dotted, solid, and dashed lines correspond to launch mass limits of 3600, 9800, and 25000 kg, respectively.  The color of each plotted point corresponds to the starshade diameter as shown by the color bar on the right.  Next to each data point we also list the optimum number of stars and slews chosen by the code, given in the format $n_{\rm slews}$/$n_{\rm targets}$.  Note that the number of slews is equivalent to the number of observations.  Observations with zero slews correspond to dry masses that exceeded the payload limits.

In each plot, we report the sensitivity of the yield to changes in the relevant mission parameters, $\phi$.  For each launch mass limit, we provide $\phi$ at the baseline values, with each value of $\phi$ evaluated at the corresponding symbol on the curve.  For example, the left panel shows that at the location of the downward pointed triangle, the sensitivity of the yield to changes in $D$ is $0.11$.  As discussed in \starketal\cite{stark2014_2}, the yield sensitivity is calculated as the fractional change in yield per fractional change in mission parameter, i.e.
\begin{equation}
	\phi\left(x_0\right) = \frac{\Delta N_{\rm EC}}{\Delta x} \frac{x_0}{N_{\rm EC}},
\end{equation}
where $x$ is the independent mission parameter and $N_{\rm EC}$ is the exoEarth candidate yield at $x_0$.  Note that $\phi(x_0)$ is equivalent to the exponent of the power law relationship between $N_{\rm EC}$ and $x$ at $x=x_0$, i.e., near $x=x_0$, $N_{\rm EC} \propto x^{\phi(x_0)}$.  We note that while the sensitivity metric is equivalent to the exponent of the power law relationship at the location of the red dot, the entirety of the yield curves are clearly not well described by power law relationships.

For launch mass limits of 3600 and 9800 kg, the yield has a moderate dependence on telescope aperture size.  For comparison, \starketal\cite{stark2015} showed the yield for a coronagraph scales roughly as $D^2$.  This moderate dependence results from two facts.  First, the telescope aperture impacts the photon collection rate and the PSF size, but it does not affect the IWA as it does for a coronagraph.  More importantly, the starshade does not operate entirely in the exposure time-limited regime as a coronagraph mission does, so shorter exposure times have less impact on the total yield.  Finally, as the color of the dots indicate, larger apertures require larger starshades at greater distances, which increases fuel use.  This latter factor can actually degrade the yield in some cases, as shown by the dotted line in Figure \ref{yield_vs_instrument_params}.

The left panel of Figure \ref{yield_vs_instrument_params} also tells us about the ideal ratio of $n_{\rm slews}$/$n_{\rm targets}$.  For small $D$, where the mission is operating closer to the exposure time-limited regime, $n_{\rm slews} > n_{\rm targets}$, indicating that revisits are valuable.  As the aperture size increases, exposure times decrease and slews become more expensive, such that the mission is operating closer to the fuel-limited regime.  In the fuel-limited regime, $n_{\rm slews} \sim n_{\rm targets}$, indicating that (potentially longer) single visits are more productive.

The middle panel in Figure \ref{yield_vs_instrument_params} show the yield as a function of IWA for the baseline mission.  An optimum IWA is clearly evident.  The yield decreases at small IWA because large, massive starshades are required with large separation distances, greatly impacting the number of slews that can be made.  On the other hand, the yield is reduced at large IWA because fewer HZs are accessible.  As previously discussed, in the fuel-limited regime of small IWA, $n_{\rm slews} \sim n_{\rm targets}$.  In the large IWA scenario, $n_{\rm slews} \approx 2n_{\rm targets}$, indicating that revisits are very valuable, a result of operating closer to the exposure time-limited regime and having fewer accessible HZs.  We note that the optimum IWA appears to vary with the launch mass limit while all other parameters are kept fixed.  

As the right panel in Figure \ref{yield_vs_instrument_params} illustrates, the yield is generally a weak function of contrast.  As long as the starshade mass does not dominate the launch mass limit, yield generally decreases as contrast is degraded.  This is in spite of the fact that as contrast is degraded, starshade diameter and separation are reduced (as indicated by the color of the data points).

\begin{figure}[H]
\centering
\includegraphics[width=6.5in]{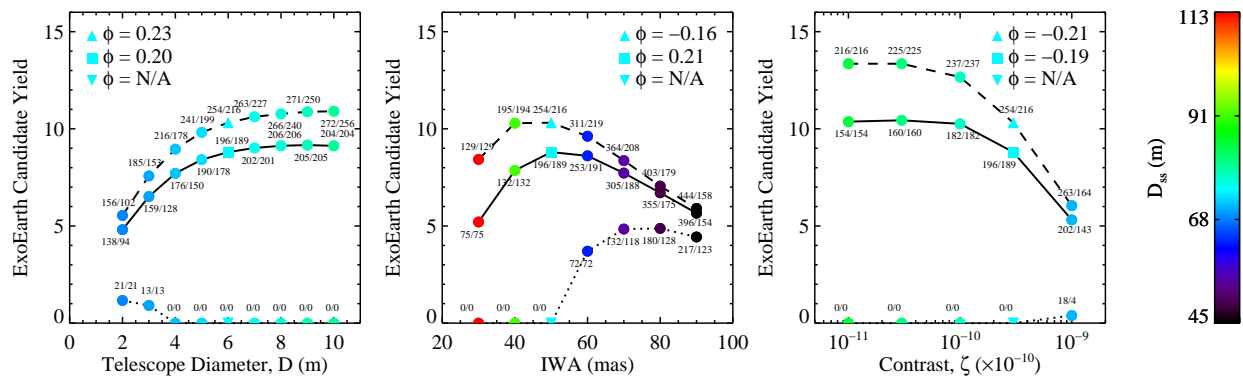}
\caption{Same as Figure \ref{yield_vs_instrument_params}, but for values of $D = 6$ m, IWA$ = 50$ mas, and $\zeta = 3\times10^{-10}$.\label{yield_vs_instrument_params2}}
\end{figure}

The sensitivity of the yield to changes in mission parameters varies significantly over parameter space.  For example, Figure \ref{yield_vs_instrument_params2} shows the same sensitivity curves as Figure \ref{yield_vs_instrument_params} when adopting baseline values of $D = 6$ m, IWA$ = 50$ mas, and $\zeta = 3\times10^{-10}$.  However, the qualitative shape of the yield curves remain similar.

\begin{figure}[H]
\centering
\includegraphics[width=6.5in]{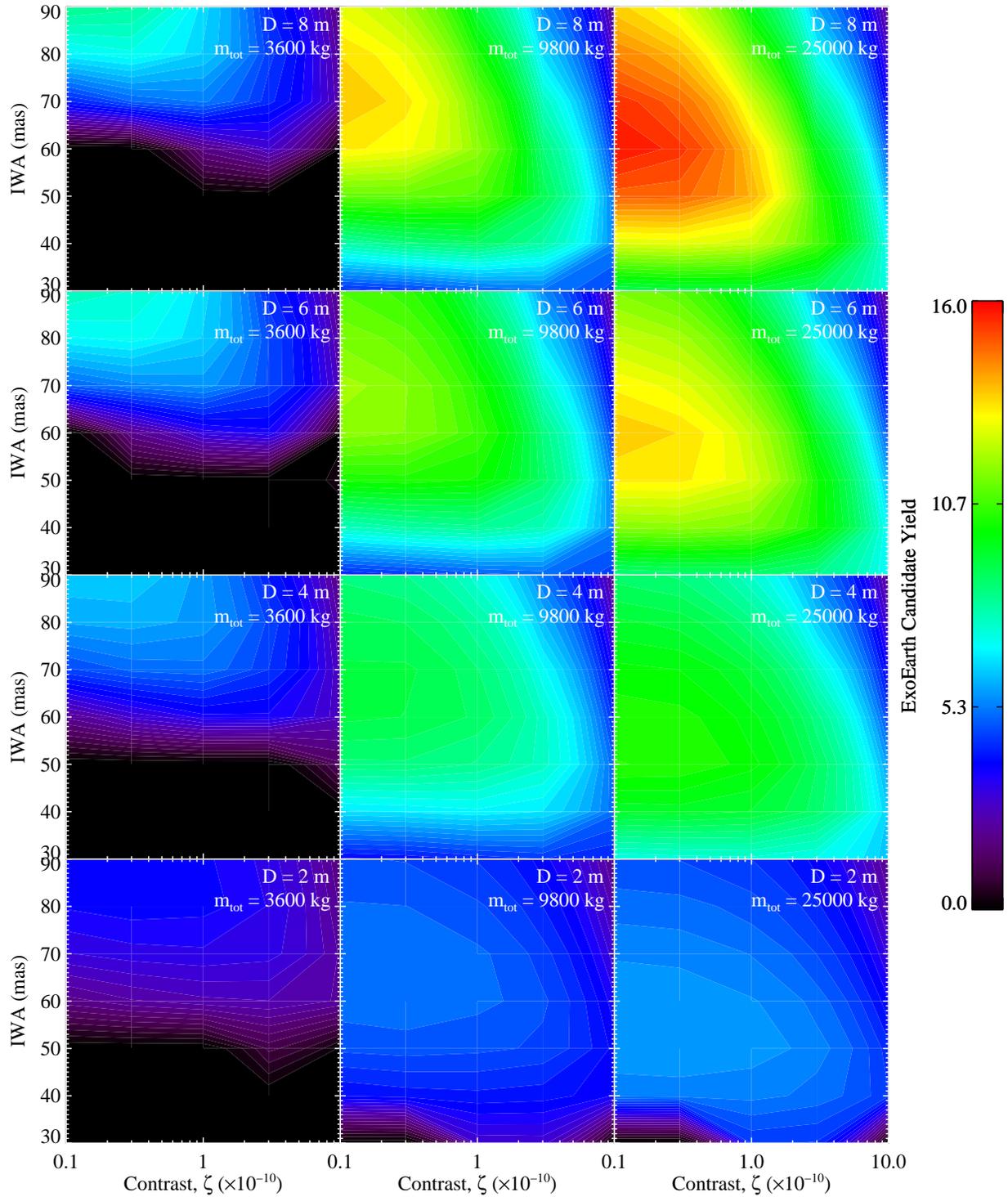}
\caption{ExoEarth candidate yield contours as a function of IWA and contrast.  Rows correspond to aperture sizes ranging from 2--8 m and columns correspond to total initial launch masses ranging from 3600--25000 kg.\label{yield_surface}}
\end{figure}

Figure \ref{yield_surface} shows the exoEarth candidate yield over 4 dimensions.  We plot 2D contours of yield as a function of IWA and contrast for 4 different telescope apertures (rows) and 3 different launch mass limits (columns).  The lower-left corner of each plot shows reduced yield---this region correspond to high-contrast and small IWA such that the starshade payload mass can dominate the launch mass limit, leaving little room for fuel.  An optimal IWA clearly exists, and it does not appear to be a strong function of $D$; the launch mass limit appears to predominantly determine the optimal starshade IWA.

\begin{figure}[H]
\centering
\includegraphics[width=6in]{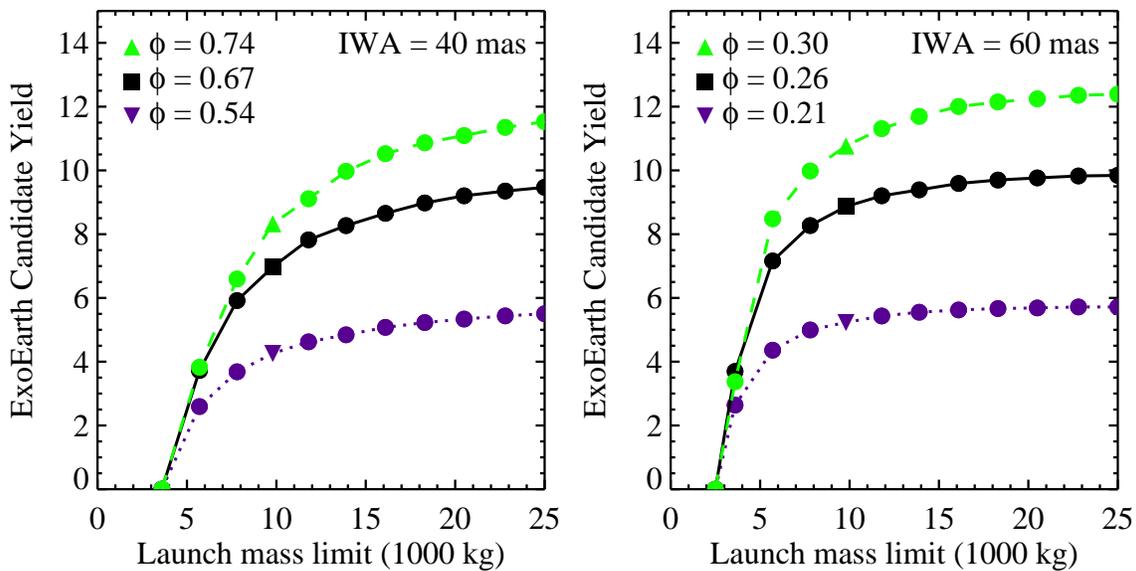}
\caption{ExoEarth candidate yield as a function of total launch mass.  Lines correspond to telescope apertures of 2 (purple dotted), 4 (black solid), and 6 m (green dashed).  \label{yield_vs_mtot}}
\end{figure}

Figure \ref{yield_vs_mtot} shows exoEarth candidate yield as a function of launch mass limit for two different IWAs and three different telescope apertures. The purple dotted, black solid, and green dashed lines correspond to telescope apertures of 2, 4, and 6 m, respectively.  Larger launch mass limits correspond to larger initial fuel masses.  In all cases, the yield ramps up quickly at low mass limits, then flattens out such that yield becomes a weak function of initial fuel mass.  This flattening of the yield curves simply reflects the exponential nature of the rocket mass equation; to achieve an additional slew we must add fuel, but that additional fuel adds mass that we must move.

\begin{figure}[H]
\centering
\includegraphics[width=4in]{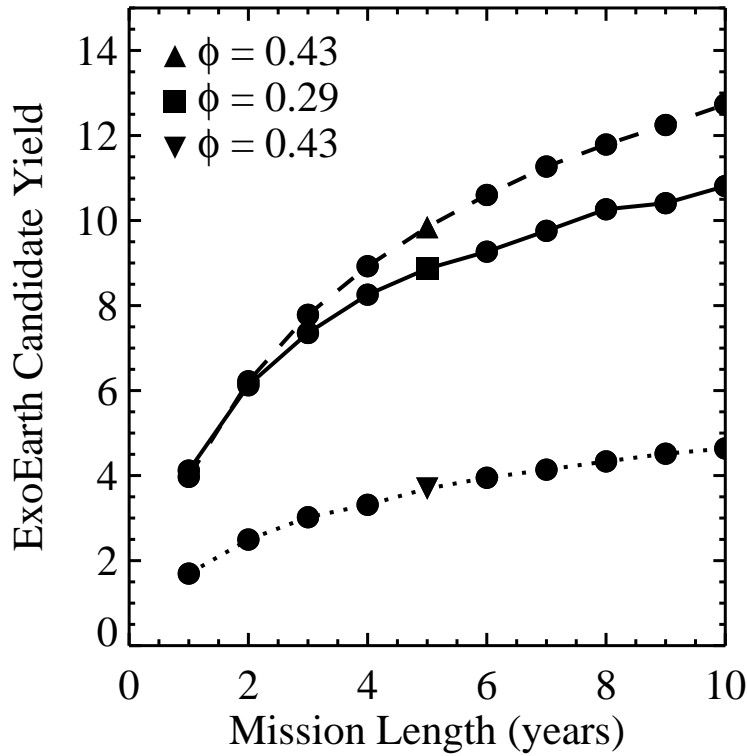}
\caption{ExoEarth candidate yield as a function of mission length for launch mass limits of 3600 (dotted line), 9800 (solid line), and $25000$ kg (dashed line).  \label{yield_vs_mission_lifetime}}
\end{figure}

Figure \ref{yield_vs_mission_lifetime} shows exoEarth candidate yields as a function of total mission length for the baseline mission, where launch mass limits of 3600, 9800, and 25000 kg are indicated by dotted, solid, and dashed lines, respectively. Although yield is a moderately weak function of total mission length, it is on par with a coronagraph's relationship between yield and total exposure time \cite{stark2015}.  This result may be somewhat surprising, as starshades do not operate entirely in the exposure time-limited regime and yield curves shown above have been relatively insensitive to the photon collection rate.  However, there is a difference between exposure time (photon collection rate) and total mission time, and Figure \ref{yield_vs_mission_lifetime} shows the latter, which can be used to lengthen slew time and reduce fuel consumption per slew.

\begin{figure}[H]
\centering
\includegraphics[width=6.5in]{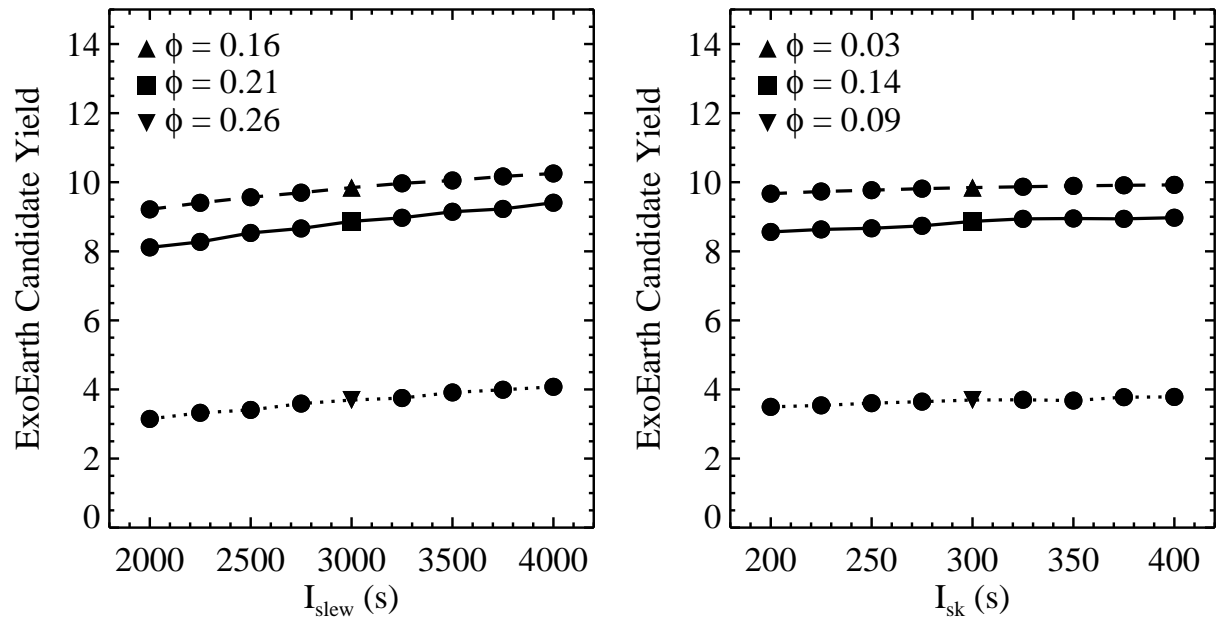}
\caption{ExoEarth candidate yield as a function of propulsion parameters $I_{\rm slew}$ (left) and $I_{\rm sk}$ (right). Dotted, solid, and dashed lines correspond to launch mass limits of 3600, 9800, and $25000$ kg, respectively.\label{yield_vs_Isp}}
\end{figure}

Because the yield of a starshade depends on future propulsion technology, the sensitivity of the yield to propulsion parameters is an important check on starshade performance.  The left and right panels of Figure \ref{yield_vs_Isp} show exoEarth candidate yield as a function of $I_{\rm slew}$ and $I_{\rm sk}$, respectively, for each of the assumed launch mass limits.  The yield for our baseline starshade mission is weakly dependent on the specific impulse for both slewing and station-keeping; a factor of 2 change in the specific impulse (or any other factor in the slew fuel use expression) impacts the baseline yield by $\sim15\%$.  We note that these scaling relationships are not the result of operating far from the thrust-limited regime (we find similar sensitivity with $\mathcal{T} = 1$ N), but will change depending on the location in parameter space.  For example, a smaller aperture paired with a more massive starshade payload would lead to a greater dependence on specific impulse.

\begin{figure}[H]
\centering
\includegraphics[width=4in]{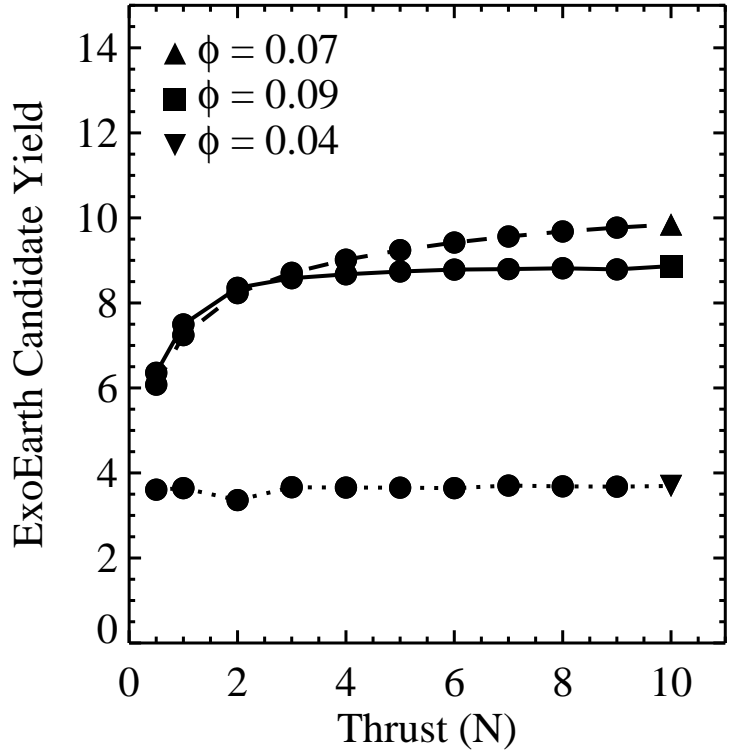}
\caption{ExoEarth candidate yield as a function of thrust. Dotted, solid, and dashed lines correspond to launch mass limits of 3600, 9800, and $25000$ kg, respectively.\label{yield_vs_thrust}}
\end{figure}

All of the calculations in this paper assume a thrust of 10 N such that the starshade is operating far from the thrust-limited regime.  To illustrate this regime, Figure \ref{yield_vs_thrust} shows the yield as a function of thrust.  For a launch mass limit of 3600 kg, the yield is flat for $\mathcal{T} > 0.5$ N.  For launch mass limits of 9800 and 25000 kg, the yield is thrust limited for $\mathcal{T} \lesssim 2$ N and $\mathcal{T} \lesssim 8$ N, respectively; our adopted thrust of 10 N ensures that all simulations do not operate in the thrust-limited regime.  More realistic, lower thrust values $\sim1$ N could reduce the baseline mission yields by $\sim15\%$.

\begin{figure}[H]
\centering
\includegraphics[width=6.5in]{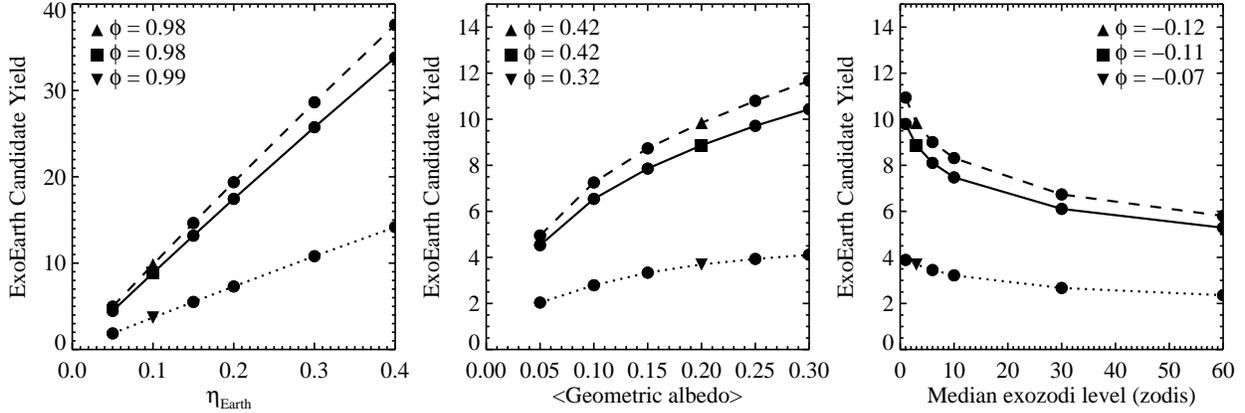}
\caption{ExoEarth candidate yield as a function of astrophysical parameters for launch mass limits of 3600 (dotted line), 9800 (solid line), and $25000$ kg (dashed line).  \label{yield_vs_astrophysical_params}}
\end{figure}

Finally, Figure \ref{yield_vs_astrophysical_params} shows the yield of our baseline mission as a function of the astrophysical assumptions $\eta_{\rm Earth}$, average exoEarth candidate albedo, and median exozodi level for the three different launch mass limits.  Comparing Figure \ref{yield_vs_astrophysical_params} to Figure 14 of \starketal\cite{stark2015}, we see that our baseline starshade is less sensitive to both the average geometric albedo of exoEarth candidates and exozodi level, consistent with the starshade not operating in the exposure time-limited regime.

\section{Discussion}
\label{discussion_section}

The scaling relationships for a starshade mission's exoEarth candidate yield are much more variable than those for a coronagraph, and depend greatly upon the specific set of mission parameters being considered---specifically, the degree to which the starshade is operating in the fuel- or exposure-time limited regime.  That being said, there are several obvious conclusions given the yield curves presented above.

First, starshades do not have any mission parameters that can impact yield to the same degree that aperture size impacts a coronagraph's yield.  Although starshade yield can be extremely sensitive to IWA ($\propto$IWA$^{4.3}$), this is limited to isolated regions of parameter space in which the starshade design is far from the optimal IWA.  The optimal IWA limits the yield gain that can be realized from this relationship.

Assuming a mission adopts an IWA close to the optimum IWA, we found yield could be as sensitive to aperture size as $\propto D^{0.6}$, but the exponent depends significantly on the mission parameters and is generally $<0.6$.  However, this relationship also cannot increase yield without bound, and the yield curves eventually turn over at larger aperture sizes as starshade diameter, distance, and mass become too great.  Because of this modest dependance on aperture size, the yield of a starshade mission with a 4 m aperture is only $\sim20$--$50\%$ greater than that of a $2.4$ m aperture, assuming a fixed IWA and contrast.

The yield of a starshade mission may be most sensitive to a parameter we did not consider: the number of starshades.  Given the exponential nature of the rocket mass equation and the yield curves shown in Figure \ref{yield_vs_mtot}, it is clearly more advantageous to launch 2 starshades with half the total launch mass than 1 starshade with the full launch mass limit, if possible.  We leave detailed calculations of multiple starshades to future work, but speculate that doubling the number of starshades should produce slightly less than a doubling of yield as long as the two starshade payloads do not dominate the total launch mass limit.  For example, whereas our baseline yield shown by the black square in the right panel of Figure \ref{yield_vs_mtot} achieved $\sim9$ exoEarth candidates, two such starshades may achieve $\sim15$ exoEarth candidates.

Under our baseline astrophysical assumptions, the largest yield estimate we calculated is $\approx16$ exoEarth candidates.  We were unable to find a set of mission parameters that produced a yield of $\sim$30--60 exoEarth candidates, the yield goals suggested by \starketal\cite{stark2014_2,stark2015}.  As shown in Figure \ref{yield_vs_astrophysical_params}, the absolute yield of a starshade mission of course depends strongly on the value of $\eta_{\rm Earth}$.  If our best estimate of $\eta_{\rm Earth}$ increases by a factor of a few, a starshade mission could conceivably achieve such yields.  We note that for consistency, we adopted $\eta_{\rm Earth}=0.1$, as was used in \starketal\cite{stark2015}.  \starketal\cite{stark2015} adopted the 1$\sigma$ lower limit on $\eta_{\rm Earth}$ from recent planet occurrence rate estimates, explicitly accounting for uncertainty in this important parameter.  The expected $\eta_{\rm Earth}$ value will likely evolve with time as our estimates improve, though it is unclear how or whether the 1$\sigma$ lower limit on $\eta_{\rm Earth}$ will change.

Assuming the fuel mass scaling relationships presented in this paper hold true when performing detailed dynamical calculations for the starshade and telescope orbits, the yield curves shown in this paper may be optimistic.  We assumed a very large value of thrust (10 N), specifically chosen to operate far from the thrust-limited regime; real starshades may indeed be thrust-limited.  We also did not budget for telescope mass and explicitly assumed the telescope and starshade launched separately.

Although a mission with a single starshade appears unable to achieve the same yield as a coronagraph mission under our baseline astrophysical assumptions, it's interesting to note that the starshade is less sensitive to photometric noise sources as well as parameters controlling the planet's photon collection rate.  As shown by Figure \ref{yield_vs_astrophysical_params}, the yield is a weak function of average exoEarth candidate albedo and a very weak function of median exozodi level.  Starshades are more robust to some astrophysical sources of uncertainty.

Finally, we discuss two assumptions made by \starketal\cite{stark2015} that become important when comparing the performance of starshades to coronagraphs.  First, \starketal\cite{stark2015} assumed, as we do in this paper as well, that the spectral characterization time required to achieve an $R=50$, S/N$=5$ spectrum is only devoted to planets that are actually exoEarth candidates (Earth-sized planets in the HZ).  This implicitly assumes that we can differentiate between exoEarth candidates and other sources (background objects, crescent phase giant planets, etc.) via a combination of color information, debris disk/orbital inclination, absolute brightness, etc., and that we can obtain this information without affecting the initial detection exposure time (this implies that we may require simultaneous multi-band imaging)\footnote{This may sound like no time can be ``wasted" on background objects or other types of planets, an optimistic assumption.  However, because of the statistical nature of the detections and the most probable phase of finding an exoEarth, if one performs the detections with an energy-resolving detector, we will likely obtain some spectra for free, as shown by the distribution of realized S/N in Figure 13 of \starketal\cite{stark2015}.}.  This assumption may be more valid for a starshade given the continuous spectral coverage from $0.5$--$1$ $\mu$m, and the negative impacts of such an assumption on yield would likely be less for the starshade than coronagraph since it does not operate in the exposure time-limited regime.

Second, \starketal\cite{stark2015} assumed, as we do in this paper as well, that no revisits are performed with the intent of establishing an orbit.  If revisits are required to establish a HZ orbit for a given target, then the yield of the fuel-limited starshade would be impacted more significantly than that of the coronagraph.  Assuming a starshade mission can perform $\sim150$ slews, if each exoEarth candidate requires $\sim5$ observations for orbit determination, a substantial amount of fuel would be devoted to such follow-up.  The coronagraph, on the other hand, would either devote some of its search time to such follow-up (which may not significantly impact yield as the yield of a coronagraph mission is a moderately weak function of total exposure time), or could simply perform such observations during an extended mission.

\section{Conclusions}
\label{conclusions}

We presented analytic expressions for starshade fuel use and combined these with our yield optimization methods to estimate maximized exoEarth candidate yields for starshade-based missions, adopting the same astrophysical assumptions as \starketal\cite{stark2015} did for coronagraph-based missions.  We find that starshade yields are maximized when the starshade operates somewhere between the time- and fuel-limited regimes.  As a result, starshade yields are less sensitive to astrophysical photometric noise sources (e.g., exozodi and unsuppressed starlight) and average exoEarth albedo; starshades are more robust to some astrophysical uncertainties, with the notable exception of $\eta_{\Earth}$.  However, this also results in starshade yields that are less sensitive to mission parameters controlling the planet photon collection rate (e.g., telescope aperture).  Under the assumption $\eta_{\Earth} = 0.1$, we are unable to find a set of mission parameters that provides a yield of several dozen exoEarth candidates for a 5 year mission with a single starshade.  If $\eta_{\Earth} = 0.3$, a yield of several dozen becomes possible, but only with a telescope aperture $\gtrsim 7$ m, a $\sim 70$ m diameter starshade, a thrust of $\gtrsim2$ N, and the launch mass limits of a Delta IV Heavy dedicated solely to the starshade.  If $\eta_{\Earth} \gtrsim 0.4$, a yield of several dozen could be achieved with a telescope aperture $\sim4$ m.  Flying multiple starshades, which we do not model in this paper, would relax these requirements.

\acknowledgments

The results reported herein benefitted from collaborations and/or information exchange within NASA's Nexus for Exoplanet System Science (NExSS) research coordination network sponsored by NASA's Science Mission Directorate.  A.R. and A.M.M. acknowledge support by GSFC's internal research and development fund.


\bibliography{ms_v3.bbl}

\begin{thebibliography}{10}

\bibitem{brown2005}
R.~A. {Brown}, ``{Single-Visit Photometric and Obscurational Completeness},''
  {\em \apj} {\bf 624}, 1010--1024  (2005).

\bibitem{brownsoummer2010}
R.~A. {Brown} and R.~{Soummer}, ``{New Completeness Methods for Estimating
  Exoplanet Discoveries by Direct Detection},'' {\em \apj} {\bf 715}, 122--131
  (2010).

\bibitem{stark2014_2}
C.~C. {Stark}, A.~{Roberge}, A.~{Mandell}, and T.~D. {Robinson}, ``{Maximizing
  the ExoEarth Candidate Yield from a Future Direct Imaging Mission},'' {\em
  \apj} {\bf 795}, 122  (2014).

\bibitem{stark2015}
C.~C. {Stark}, A.~{Roberge}, A.~{Mandell}, M.~{Clampin}, S.~D.
  {Domagal-Goldman}, M.~W. {McElwain}, and K.~R. {Stapelfeldt}, ``{Lower Limits
  on Aperture Size for an ExoEarth Detecting Coronagraphic Mission},'' {\em
  \apj} {\bf 808}, 149  (2015).

\bibitem{cash2006}
W.~{Cash}, ``{Detection of Earth-like planets around nearby stars using a
  petal-shaped occulter},'' {\em \nat} {\bf 442}, 51--53  (2006).

\bibitem{vanderbei2007}
R.~J. Vanderbei, E.~Cady, and N.~J. Kasdin, ``Optimal occulter design for
  finding extrasolar planets,'' {\em The Astrophysical Journal} {\bf 665}(1),
  794  (2007).

\bibitem{kopparapu2013}
R.~K. {Kopparapu}, R.~{Ramirez}, J.~F. {Kasting}, V.~{Eymet}, T.~D. {Robinson},
  S.~{Mahadevan}, R.~C. {Terrien}, S.~{Domagal-Goldman}, V.~{Meadows}, and
  R.~{Deshpande}, ``{Habitable Zones around Main-sequence Stars: New
  Estimates},'' {\em \apj} {\bf 765}, 131  (2013).

\bibitem{leinert1998}
C.~{Leinert}, S.~{Bowyer}, L.~K. {Haikala}, M.~S. {Hanner}, M.~G. {Hauser},
  A.-C. {Levasseur-Regourd}, I.~{Mann}, K.~{Mattila}, W.~T. {Reach},
  W.~{Schlosser}, H.~J. {Staude}, G.~N. {Toller}, J.~L. {Weiland}, J.~L.
  {Weinberg}, and A.~N. {Witt}, ``{The 1997 reference of diffuse night sky
  brightness},'' {\em \aaps} {\bf 127}, 1--99  (1998).

\bibitem{exos2015}
S.~{Seager}, W.~{Cash}, S.~{Domagal-Goldman}, N.~J. {Kasdin}, M.~{Kuchner},
  A.~{Roberge}, S.~{Shaklan}, W.~{Sparks}, M.~{Thomson}, M.~{Turnbull},
  K.~{Warfield}, D.~{Lisman}, R.~{Baran}, R.~{Bauman}, E.~{Cady},
  C.~{Heneghan}, S.~{Martin}, D.~{Scharf}, R.~{Trabert}, D.~{Webb}, and
  P.~{Zarifian}, ``{Exo-S: Starshade Probe-Class Exoplanet Direct Imaging
  Mission Concept Final Report},'' {\em NASA ExoPlanet Exploration Program, Jet
  Propulsion Laboratory, California Institute of Technology}   (2015).

\bibitem{rioux2015}
N.~Rioux, H.~Thronson, L.~Feinberg, H.~P. Stahl, D.~Redding, A.~Jones,
  J.~Sturm, C.~Collins, and A.~Liu {\em Proc. SPIE} {\bf 9602},
  960205--960205--12  (2015).

\bibitem{cady2009}
E.~Cady, K.~Balasubramanian, M.~Carr, M.~Dickie, P.~Echternach, T.~Groff,
  J.~Kasdin, C.~Laftchiev, M.~McElwain, D.~Sirbu, R.~Vanderbei, and V.~White,
  ``Progress on the occulter experiment at princeton,'' {\em Proc. SPIE} {\bf
  7440}, 744006--744006--10  (2009).

\bibitem{cady2011}
E.~Cady, ``Nondimensional representations for occulter design and performance
  evaluation,'' {\em Proc. SPIE} {\bf 8151}, 815112--815112--10  (2011).

\bibitem{kolemenkasdin2007}
E.~Kolemen and N.~J. Kasdin, ``Optimal trajectory control of an occulter based
  planet finding telescope,'' {\em Proceedings of the AAS Guidance and Control
  Conference at Breckenridge, Colorado}   (2007).
\newblock AAS 07-037.

\bibitem{savransky2010}
D.~{Savransky}, N.~J. {Kasdin}, and E.~{Cady}, ``{Analyzing the Designs of
  Planet-Finding Missions},'' {\em \pasp} {\bf 122}, 401--419  (2010).

\bibitem{hunyadi2007}
S.~L. {Hunyadi}, S.~B. {Shaklan}, and R.~A. {Brown}, ``{The lighter side of
  TPF-C: evaluating the scientific gain from a smaller mission concept},'' in
  {\em Society of Photo-Optical Instrumentation Engineers (SPIE) Conference
  Series},  {\em Society of Photo-Optical Instrumentation Engineers (SPIE)
  Conference Series} {\bf 6693}  (2007).

\bibitem{turnbull2012}
M.~C. {Turnbull}, T.~{Glassman}, A.~{Roberge}, W.~{Cash}, C.~{Noecker},
  A.~{Lo}, B.~{Mason}, P.~{Oakley}, and J.~{Bally}, ``{The Search for Habitable
  Worlds. 1. The Viability of a Starshade Mission},'' {\em \pasp} {\bf 124},
  418--447  (2012).

\bibitem{trabert2015}
R.~Trabert, S.~Shaklan, P.~D. Lisman, A.~Roberge, M.~Turnbull,
  S.~Domagal-Goldman, and C.~Stark, ``Design reference missions for the
  exoplanet starshade (exo-s) probe-class study,'' {\em Proc. SPIE} {\bf 9605},
  96050Y--96050Y--12  (2015).

\bibitem{seager2015}
S.~Seager, M.~Turnbull, W.~Sparks, M.~Thomson, S.~B. Shaklan, A.~Roberge,
  M.~Kuchner, N.~J. Kasdin, S.~Domagal-Goldman, W.~Cash, K.~Warfield,
  D.~Lisman, D.~Scharf, D.~Webb, R.~Trabert, S.~Martin, E.~Cady, and
  C.~Heneghan, ``The exo-s probe class starshade mission,'' {\em Proc. SPIE}
  {\bf 9605}, 96050W--96050W--18  (2015).

\end{thebibliography}



\end{spacing}
\end{document}